\documentclass[10pt,preprint2]{aastex6}
\usepackage[utf8x]{inputenc}
\usepackage{amsmath}
\usepackage{amsfonts}
\usepackage{amssymb}
\usepackage{mathrsfs}
\usepackage{bm}

\usepackage{CJK}
\usepackage{ucs}
\newcommand{\cntext}[1]{\begin{CJK}{UTF8}{bsmi}#1\end{CJK}}

\defcitealias{1992ApJ...389..129A}{AL92}
\defcitealias{2015ApJ...815L..21F}{FD15}
\defcitealias{2002MNRAS.330..950O}{OL02}

\newcommand{\GG}[1]{}

\begin{document}
\title{Ultraharmonics and Secondary Spiral Wakes Induced by a Planet}
\author{Wing-Kit Lee \cntext{(李詠傑)}}
\affil{Institute of Astronomy and Astrophysics, Academia Sinica\\P.O. Box 23-141, Taipei 10617, Taiwan}
\email{wklee@asiaa.sinica.edu.tw}
\shorttitle{SECONDARY SPIRALS}
\shortauthors{LEE}

\begin{abstract}
We investigate the ultraharmonics response of a protoplanetary disk to an orbiting planet. We find that the multi-armed spiral structure can be excited by the higher-order forcing due to nonlinear mode-coupling. In particular, the preferential excitation of gas response with small azimuthal wavenumber $(m\sim 2)$ is a direct consequence of mode-coupling among linear waves. The presence of multiple Fourier components in a planet's potential is a distinct feature compared to the previous studies in the context of spiral galaxies, which turns out to be crucial for the generation of ultraharmonics waves. This analysis may shed light on understanding some results regarding the spiral structures excited by a massive planet.
\end{abstract}

\section{Introduction}

\begin{figure}[!htb]
\centering
\includegraphics[width=0.5\textwidth]{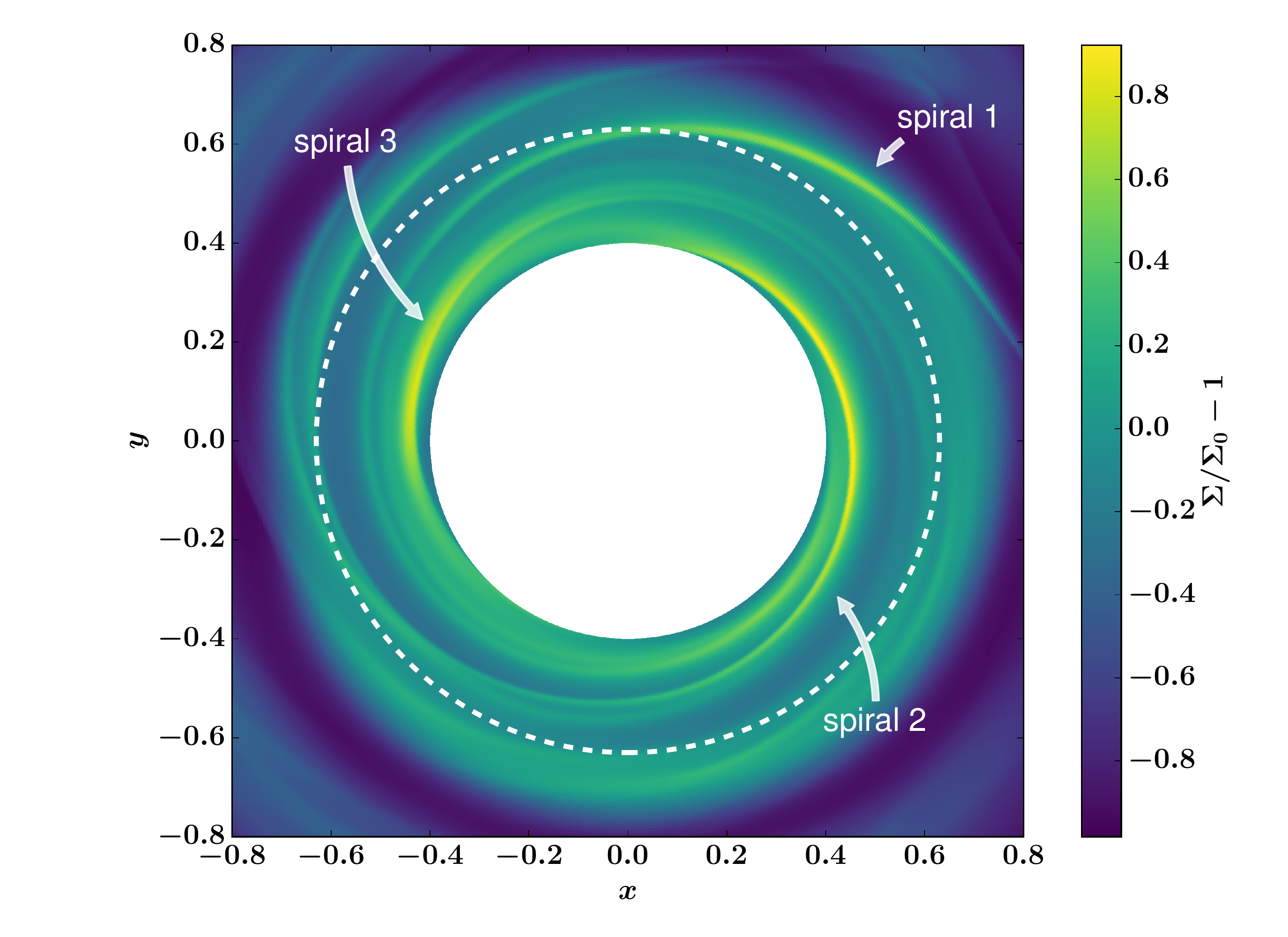}
\caption{A typical gas response due to a Jupiter-mass planet ($q=10^{-3}$) in the inner disk. The linear color scale shows the surface density perturbation relative to the equilibrium disk, i.e., $\Sigma/\Sigma_0 - 1$. Only the inner disk is shown to emphasize the existence of multiple spiral arms. The planet is located at $(x,y)=(1,0)$ (not shown). The dashed white circle marks the LR for $m=2$. In addition to the spiral arm that connects to the planet (spiral 1), at least two more spirals can be identified. See Section \ref{sec:Section6} for numerical details.}
\label{fig0}
\end{figure}

Spiral structure in an accretion disk has been studied extensively in the literature due to its ubiquitous nature in astrophysics \citep[see a recent review by][]{ShuARAA2016}. In particular, the spirals found in protoplanetary disks may have implication to the search for exoplanets. The spiral waves can be excited due to the tidal interaction between the embedded planet and the disk \citep{1979ApJ...233..857G,1980ApJ...241..425G}, which in turn facilitates angular momentum transport across the disk \citep[e.g.,][]{1972MNRAS.157....1L,2016arXiv160103009R}, and leads to planet migration \citep[e.g.,][]{1979MNRAS.186..799L,1997Icar..126..261W,2002ApJ...565.1257T}. The departure from linear theories, such as existence of secondary spirals (as opposed to the primary spiral that traces the planet's position, see Figure \ref{fig0}), indicates the importance of nonlinear effects. One of the most relevant nonlinear effects related to additional spirals is ultraharmonics, which is a consequence of nonlinear mode-coupling. Ultraharmonics was first studied in the context of spiral galaxies, which results in a multi-armed gas response (i.e., additional branches between spiral arms) to an underlying bisymmetric stellar potential
\citep{SMR1973,1984ApJ...281..600Y,1991ApJ...376..104Y,2003ApJ...596..220C,2015ApJ...800..106W}. \citet[][hereafter AL92]{1992ApJ...389..129A} studied the excitation of ultraharmonics waves by a single $m$-fold symmetric potential that is present in a spiral galaxy. An Eulerian approach to nonlinear mode-coupling had also been used for eccentric instability \citep{1991ApJ...381..259L} and gravitational instability\footnote{It was also referred as ``restricted three-wave effect" for the two-armed spiral waves caused by self-gravity.} \citep{1996ApJ...460..855L,1998ApJ...504..945L} in accretion disks. Non-axisymmetric perturbations can also lead to higher-order correction to the mass flux \citep{1990ApJ...362..395L}. On the other hand, a Lagrangian formulation had been applied to Saturn rings \citep{1985ApJ...291..356S} and protostellar disk for a single harmonics of planet's potential \citep{1994ApJ...437..338Y}. In addition, other nonlinear effects are also present, such as gap-opening \citep[e.g.,][]{1979MNRAS.186..799L,1993prpl.conf..749L,1996ApJ...460..832T}.

The existence of spiral structure in protoplanetary disks have been supported by many near-IR observations \citep[e.g.,][]{2012ApJ...748L..22M,2013ApJ...762...48G,2015A&A...578L...6B, 2015ApJ...813L...2W,2016arXiv160300481S}. However, previous attempts to relate observed spirals to theoretical predictions of planet-induced spiral faced several challenges. One reason is that the scattered light intensity do not trace the gas surface density directly \citep{2014ApJ...795...71T, 2015MNRAS.451.1147J}. On the other hand, the inferred scale-height from the current theories based on the nonlinear steepening of spiral waves \citep{2001ApJ...552..793G,2002ApJ...572..566R} is generally too large, which implies an unphysical hot disk \citep{2012ApJ...748L..22M,2015MNRAS.451.1147J}. Interestingly, not even the number of planets can be easily constrained by the number of spiral arms \citep[e.g.,][]{2015A&A...578L...6B}. A two-armed spiral structure in the disk can be excited by two low-mass planets (one spiral each) or excited by one massive planet at a large radius \citep{2015ApJ...813...88Z,2015ApJ...809L...5D}. The whole picture is further complicated by other possible mechanisms, such as gravitational instability \citep[e.g.,][]{1998ApJ...504..945L,2015MNRAS.454.2529M}, accretion from envelope \citep[e.g.,][]{2015A&A...582L...9L}, encounter of substellar objects \citep[e.g.,][]{2015MNRAS.449.1996D, 2015ApJ...806L..10D}, shadows cast by an tilted inner disk \citep{2016arXiv160107912M}, etc.

Whether one single planet can excite multiple spiral arms and what is the physics behind are the main questions we try to address in this paper. These secondary spirals were found even in some early simulations \citep[e.g.,][]{1999MNRAS.303..696K}, although their origin and implication were not deeply discussed until recent numerical studies (\citealt{2015MNRAS.451.1147J,2015ApJ...813...88Z}; \citetalias{2015ApJ...815L..21F}). In particular, \citetalias{2015ApJ...815L..21F} found a power-law relation between angular separation of spirals and planet's mass, which has a potential application to constrain observations. The angular separation increases up to $180^{\rm o}$ for a very massive planet, which is consistent to the other studies of stellar binaries. The dependence of gas morphology on the planet's mass leads to the hypothesis that it is a nonlinear effect, in which forcing amplitude matters. In general, the linearity is determined by mass ratio between the planet and the central star, $q=M_p/M_s$. When $q \gtrsim 10^{-4}$, a secondary, weaker spiral arm is often found \citep{2006MNRAS.370..529D,2015ApJ...813...88Z} (see also, Figure \ref{fig0}). Hence, a bisymmetric spiral structure can be caused by a massive planet, such as the cases of SAO 206462 and MWC 758 where a planet with a few Jupiter mass may be present \citep{2015ApJ...813...88Z,2015ApJ...809L...5D}. 

In this paper, we formulate a theory to study the secondary spiral arms by studying the disk-planet interaction in the slightly nonlinear regime. In particular, we investigate the mechanism to produce ultraharmonics. Hence, our work is based on \citetalias{1992ApJ...389..129A} but extends to the nonlinear mode-coupling among different harmonics of the underlying planet's potential. We show that such extension is crucial to explain why low-$m$ modes such as $m=2$ are favored by a massive planet. In Section \ref{sec:Section2}, we derive the governing equations for the higher-order ultraharmonics. In Section \ref{sec:Section3}, we study a simple case of a single Fourier component of a planet's potential. The full planet's potential case is presented in Section \ref{sec:Section4}. In Section \ref{sec:force}, we study the nonlinear driving of individual ultraharmonics mode. In Section \ref{sec:Section6}, we present numerical results for different cases. Lastly, we discuss our results in Section \ref{sec:Section7} and conclude in Section \ref{sec:Section8}.


\section{Governing Equations}
\label{sec:Section2}

We first describe the basic equations of an embedded planet orbiting a central star in a cold Keplerian disk. The gas response due to the tidal interaction between a planet and a disk is studied. For simplicity, we assume a circular orbit for the planet and only consider the two-dimensional vertically-integrated gas dynamics. We consider a cylindrical coordinate system $(r,\varphi)$ that is centered at the star of mass $M_s$ and is rotating at the planet's orbital frequency $\Omega_p$.  The basic state of the gas velocity in the rotating frame is given by ${\bm u}_0 = r[\Omega(r)-\Omega_p] \mathbf{e}_\varphi$, where the disk frequency $\Omega(r)$ is determined by the radial force balance,
\begin{align}
\label{eq:eq0}
r\Omega^2 = \frac{1}{\Sigma_0}\frac{dP_0}{dr} + \frac{d\Phi_0}{dr}.
\end{align}
Here $\Phi_0 = -GM_s/r$ is the gravitational potential of the central star, $\Sigma_0$ and $P_0$ are the equilibrium values of surface density and vertically-integrated pressure of the gas, respectively, which are both functions of $r$. In the rotating frame, the Lagrangian derivative is given by
\begin{align}
\label{eq:eq1}
\frac{D}{Dt} = \frac{\partial}{\partial t} + u \frac{\partial}{\partial r} + \left[(\Omega-\Omega_p) + \frac{v}{r}\right] \frac{\partial}{\partial \varphi},
\end{align}
where ${\bm u}_1 = u \mathbf{e}_r + v \mathbf{e}_\varphi$ is the velocity derivation from the basic state. The fluid equations can be written as
\begin{align}
\label{eq:full2_1}
\frac{D \Sigma}{D t} &+ \frac{\Sigma}{r}\left[\frac{\partial}{\partial r} ( r u ) + \frac{\partial}{\partial \varphi}v \right] = 0,
\end{align}
\begin{align}
\label{eq:full2_2}
\frac{D u}{Dt} - \frac{v^2}{r} - 2\Omega v &= -\left( \frac{1}{\Sigma}\frac{\partial P}{\partial r}-\frac{1}{\Sigma_0}\frac{dP_0}{dr}\right) -\frac{\partial \Phi_p}{\partial r},
\end{align}
\begin{align}
\label{eq:full2_3}
\frac{D v}{Dt} +\frac{u v}{r} + 2 B u &= -\frac{1}{\Sigma r}\frac{\partial P}{\partial \varphi}-\frac{1}{r}\frac{\partial \Phi_p}{\partial \varphi},
\end{align}
where $\Phi_p$ is the planet's gravitational potential, $\Sigma$ and $P$ are the gas surface density and pressure, respectively. The Oort's constant $B$ is defined by
\begin{align}
B = \frac{r}{2}\frac{d\Omega}{dr} + \Omega,
\end{align}
which equals $\Omega/4$ for a Keplerian disk. It is also convenient to define the epicyclic frequency through $\kappa^2 = 4B\Omega$. For simplicity, we ignore the small pressure correction to $B$ and $\Omega$. The planet's potential in the rotating frame can be expressed as,
\begin{align}
\Phi_p(r,\varphi) = - \frac{GM_p}{(r_0^2 + r^2 - 2r_0 r \cos\varphi)^{1/2}},
\end{align}
where the planet is located at $(r,\varphi)=(r_0,0)$. We set $r_0=1$ for simplicity. To close the equations, a locally isothermal equation of state (EOS) $P = \Sigma c^2$ is adopted such that the local sound speed $c$ is a power-law in radius,
\begin{align}
\label{eq:eos}
c(r)=c_0 (r/r_0)^{-\beta/2}
\end{align}
where $\beta>0$ and $c_0$ is a constant. In this work, we assume the disk is cold such that the disk aspect ratio at $r_0$ is small, i.e., $h_0=c_0/r_0\Omega_p \ll 1$. For simplicity, the self-gravity of the gas disk and viscosity are ignored.

\subsection{Linear Analysis}
\label{sec:linear}

Since the perturbation due to a low-mass planet is small, we may linearize the equation and obtain a solution. Here we only state some important results from previous extensive analysis \citep[e.g.,][hereafter OL02]{1979ApJ...233..857G,1980ApJ...241..425G,1993ApJ...419..155A,1993ApJ...419..166A,2002MNRAS.330..950O}
 and focus on circularly orbiting planet for simplicity. A second-order ordinary differential equation can be derived by combining Equations \eqref{eq:full2_1}-\eqref{eq:full2_3}. Ignoring the pressure correction \citep{1993ApJ...419..155A}, waves are launched at the Lindblad resonances (LRs), which are located at
\begin{align}
r_L = \left(1\pm 1/m\right)^{2/3} r_0,
\end{align}
where $m>0$ is the azimuthal wavenumber. The upper and lower signs indicate the outer LRs (OLRs) and inner LRs (ILRs), respectively. The pressure waves carry angular momentum away from the planet and wind up as they propagate (and hence trailing waves). Far away from the planet, a free wave solution can be obtained by using the WKB approximation \citepalias[c.f.,][]{2002MNRAS.330..950O}. For example, the $m$-th Fourier component of radial velocity is given by
\begin{align}
\label{eq:linear1}
u_m = U_m(r) \exp i\left[\int^r_{r_L} k_m(r) dr -\frac{\pi}{4}\right],
\end{align}
where $U_m(r)$ is the slowly-varying wave amplitude and $k_m(r)$ is the radial wavenumber. The wavenumber $k_m(r)$ satisfies the usual dispersion relation of a pressure wave in a differentially rotating disk, which reads
\begin{align}
\label{eq:linear2}
\hat{\omega}^2 = \kappa^2 + k_m^2 c^2,
\end{align}
where $\hat{\omega}=m(\Omega_p-\Omega)$ is the Doppler-shifted frequency. The radial wavenumber is real for propagating waves and is further chosen to be positive ($k_m > 0$), such that the group velocity ${\rm v}_g = \hat{\omega}/k_m$ is pointing away from the planet. The phase of $-\pi/4$ in Equation \eqref{eq:linear1} is found by matching the asymptotic solution near the LRs, which can be expressed in terms of Airy functions \citep{1984ApJ...281..600Y,1987Icar...69..157M}. For barotropic perturbations, the wave amplitude can be determined by using the conservation of wave angular momentum flux \citep{1972MNRAS.157....1L}, i.e.,
\begin{align}
\label{eq:linear3}
F_A = \frac{\pi r m}{k_m}\left(\frac{\hat{\omega}^2-\kappa^2}{\hat{\omega}^2}\right)\Sigma_0|U_m|^2.
\end{align}
Without dissipation, $F_A$ is a constant and the torque on the disk is $T_m = \pm F_A$, which equals to the rate of angular momentum carried away by the waves. However, since a locally isothermal EOS is enforced, there is a background torque due to the thermal forcing \citep{2011MNRAS.415.1445L,2015MNRAS.448.3806L}. Fortunately, a conserved quantity can still be expressed concisely, that is
\begin{align}
F_A/c^2 = \textit{const.},
\end{align}
which can be shown by considering the imaginary part of the linearized equations using the WKB anatz \citep[e.g.,][]{2002ApJ...572..566R}. On the other hand, between the $m$-th ILR and OLR, the non-wave response \citep[see, e.g.,][]{1991ApJ...381..259L} is given by 
\begin{align}
\label{eq:linear4}
u_m = \frac{i\hat{\omega}}{D}\left[\frac{d\phi_m}{dr}+\frac{2\Omega}{r(\Omega-\Omega_p)}\phi_m\right],
\end{align}
where we define $D=\kappa^2-\hat{\omega}^2$ and the terms proportional to $c^2$ is neglected in the cold disk limit \citep{1979ApJ...233..857G}. We note that the phase of the non-wave response is constant in radius. Equations \eqref{eq:linear1} and \eqref{eq:linear4} are the wave and non-wave parts of the linear solution, respectively. For the purpose of this paper, they suffice to describe the nonlinear driving of the higher-order modes.

\subsection{Higher-order Equations}
\label{sec:higherorder}
To proceed, we introduce a small parameter $\epsilon$ that is proportional to the mass ratio $q$. The flow quantities can be expressed in a power series of $\epsilon$, that is,
\begin{align}
\small
\label{eq:higher1}
u = \sum^{\infty}_{n=1} \epsilon^n u^{(n)}, \quad \text{and} \quad
v = \sum^{\infty}_{n=1} \epsilon^n v^{(n)},
\end{align}
except
\begin{align}
\Sigma = \Sigma_0 + \sum^{\infty}_{n=1} \epsilon^n \Sigma^{(n)},
\end{align}
where the basic state of surface density is included. The variables $X^{(n)}$ are real quantities. It is useful to expand the Lagrangian derivative in Equation \eqref{eq:eq1} in terms of $\epsilon$, that is, $D/Dt = \mathfrak{D}_0 + \epsilon \mathfrak{D}_1 + \epsilon^2 \mathfrak{D}_2 + \cdots$, where
\begin{subequations}
\label{eq:higher2}
\begin{align}
\mathfrak{D}_0 &= (\Omega-\Omega_p) \partial_\varphi, \\
\mathfrak{D}_1 &= u^{(1)}\partial_r+\frac{v^{(1)}}{r}\partial_\varphi, \\
\mathfrak{D}_2 &= u^{(2)}\partial_r+\frac{v^{(2)}}{r}\partial_\varphi.
\end{align}
\end{subequations}
By substituting Equation \eqref{eq:higher1} into Equations \eqref{eq:full2_1}-\eqref{eq:full2_3} and collecting the second-order quantities, we have
\begin{subequations}
\label{eq:higher3}
\begin{align}
\label{eq:higher3a}
&\mathfrak{D}_0 u^{(2)} -2\Omega v^{(2)} + c^2 \frac{\partial}{\partial r} \sigma^{(2)}= - \mathfrak{D}_1 u^{(1)} + \frac{(v^{(1)})^2}{r}  \\
\label{eq:higher3b}
&\mathfrak{D}_0 v^{(2)} +2B u^{(2)} + \frac{c^2}{r}\frac{\partial}{\partial \varphi} \sigma^{(2)} = -\mathfrak{D}_1 u^{(1)} - \frac{u^{(1)} v^{(1)}}{r} \\
\nonumber
&\mathfrak{D}_0 \sigma^{(2)} + u^{(2)} \frac{d}{dr}\left(\ln \Sigma_0\right) + \frac{1}{r} \left[ \frac{\partial}{\partial r}\left(r u^{(2)}\right) + \frac{\partial v^{(2)}}{\partial \varphi} \right] \\
\label{eq:higher3c}
&= -\frac{1}{\Sigma_0}\mathfrak{D}_1 \Sigma^{(1)} - \frac{\sigma^{(1)}}{r} \left[ \frac{\partial}{\partial r} \left(r u^{(1)}\right) + \frac{\partial v^{(1)}}{\partial \varphi} \right],
\end{align}
\end{subequations}
where $\sigma^{(2)}$ is the second-order surface density perturbation (i.e., $\sigma=\Sigma/\Sigma_0-1$). Note that the driving terms on the right-hand side of Equation \eqref{eq:higher3} can be treated as known quantities by solving the linearized equations.

Higher-order equations for $n>2$ can be obtained by the similar procedures. In particular, the coefficients on the left-hand side of Equation \eqref{eq:higher3} remain the same as they are zeroth-order. This implies that, when solving the for free WKB wave solutions, the dispersion relations for these higher-order waves are the same. 

The nonlinear mode-coupling are caused by the driving terms in the right-hand side of Equation \eqref{eq:higher3}. Without self-gravity, the nonlinear terms are quadratic in flow quantities (e.g., advection term). When expressing the linear solution into a Fourier series, the nonlinear mode-coupling allow different $m$-th harmonics to give rise to various ultraharmonics response. Before attacking the full problem with multiple $m$, we consider a simplified case in the next section where only a single $\phi_m$ is considered.

\section{Single Fourier component of the Potential}
\label{sec:Section3}

Consider a Fourier component of the planet's potential, which can be written as
\begin{align}
\label{eq:singlecomp1}
\Phi_p = \phi_m(r) \cos(m\varphi),
\end{align}
where we set $\phi_m$ to be real by fixing the phase. The solution to Equation \eqref{eq:linear1} contains only $m$-harmonics (i.e., $m$-armed response with an angular dependence $\exp{im\varphi}$). The quadratic nonlinear terms in Equation \eqref{eq:higher3} are responsible for $m'=0$ and $m'=2m$ second-order $\mathcal{O}(\epsilon^2)$ modes (of $\exp{im'\varphi}$). A second-order flow quantity $A$ can be expressed as
\begin{align}
\label{eq:singlecomp2}
A^{(2)} = \Re[ A^{(2)}_0 + A^{(2)}_2e^{i2m\varphi}],
\end{align}
where $\Re[\cdots]$ denotes the real part and $A^{(n)}_l$ is the $n$-th order response of $lm$-harmonics (of $\exp{ilm\varphi}$). Note that $A^{(2)}_1=0$ as $A^{(1)}_0=0$. Here the axisymmetric term $A^{(2)}_0$ term corresponds to a correction to the mass flux \citep{1990ApJ...362..395L}. The governing equations of the ultraharmonics response $A^{(2)}_{2}$ can be written in the complex form,
\begin{align}
\label{eq:singlecomp3}
&-il\hat{\omega} u^{(2)}_{2} - 2\Omega v^{(2)}_{2} + c^2\frac{d}{dr}\sigma^{(2)}_{2} = R^{(2)}_r, \\
\label{eq:singlecomp4}
&-il\hat{\omega} v^{(2)}_{2} +2B u^{(2)}_{2} + \frac{imlc^2}{r}\sigma^{(2)}_{2} = R^{(2)}_\varphi, \\
\nonumber
&-il\hat{\omega} \sigma^{(2)}_{2} + u^{(2)}_{2}\frac{d}{dr}\ln\Sigma_0 + \frac{1}{r}\frac{d}{dr}\left(r u^{(2)}_{2}\right) + \frac{iml}{r} v^{(2)}_{2} \\
&= R^{(2)}_c,
\label{eq:singlecomp5}
\end{align}
where $l=2$ and
\begin{subequations}
\label{eq:singlecomp6}
\begin{align}
\label{eq:singlecomp6a}
R^{(2)}_r &= -\frac{1}{2}\left(u_{m} \frac{du_{m}}{dr} + \frac{im v_{m} u_{m}}{r} - \frac{v^2_{m}}{r}\right), \\
\label{eq:singlecomp6b}
R^{(2)}_\varphi &= -\frac{1}{2}\left(u_{m} \frac{dv_{m}}{dr} + \frac{im v^2_{m}}{r} + \frac{u_{m} v_{m}}{r}\right),\\
\label{eq:singlecomp6c}
R^{(2)}_c &= - \frac{1}{2r\Sigma_0}\frac{d}{d r} (r \Sigma_{m} u_{m}) - \frac{iml}{2r\Sigma_0} \Sigma_{m} v_{m}.
\end{align}
\end{subequations}
For clarity, the superscripts $(1)$ of linear solutions in Equation \eqref{eq:singlecomp6} are dropped and we use subscript $m$ to indicate the $m$-th harmonics. The factor of $1/2$ in Equation \eqref{eq:singlecomp6} comes from the following relation
\begin{align*}
\Re(u_{m}e^{im\varphi}) \Re(v_{m}e^{im\varphi}) 
&= \Re \left[ \frac{1}{2} u_{m} v_{m}^* + \frac{1}{2} u_{m} v_{m}e^{i2m\varphi} \right],
\end{align*}
where the asterisk denotes the complex conjugate. Similar to the linear case, the resonance is apparent when we solve Equation \eqref{eq:singlecomp3}-\eqref{eq:singlecomp5}, where the denominator is given by
\begin{align}
\label{eq:singlecomp7}
D_l = \kappa^2 - l^2 m^2(\Omega_p -\Omega)^2.
\end{align}
Therefore, the ultraharmonics resonances (URs) for a $lm$-symmetric response are given by $D_l=0$. For a Keplerian disk, the URs are located at
\begin{align}
\label{eq:singlecomp8}
r_2 = \left(1 \pm 1/lm \right)^{2/3} r_0.
\end{align}
Since the URs are generally located within the region bounded by LRs of the same $m$, only the non-wave linear response is responsible (c.f., Equation \eqref{eq:linear4}). We note that the driving terms in our case is different from the one studied by \citetalias{1992ApJ...389..129A}, in which self-gravitating stellar waves propagate between LRs. We postpone the discussion of the nonlinear forcing in the next section.

\section{Full Planet's Potential}
\label{sec:Section4}

To study the full problem, we express the planet's potential as a Fourier series
\begin{align}
\label{eq:fullpot1}
\Phi_p(r,\varphi) = \sum^\infty_{m=1} \phi_m(r) \cos(m\varphi),
\end{align}
where the orbit is assumed to have zero eccentricity $e=0$ and zero inclination. With only one frequency, we can have a corotating frame for time-steady equations \citep{1980ApJ...241..425G}. To proceed, we further express the $n$-th order flow quantity $A^{(n)}$ in a Fourier series,
\begin{align}
\label{eq:fullpot2}
A^{(n)} = \Re\left[ \sum^{\infty}_{l=0} A^{(n)}_{l}(r) e^{il \varphi} \right],
\end{align}
where we denote $l$ as the azimuthal wavenumber (which differs from the definition in the Section \ref{sec:Section3}). Here the complex Fourier component is given by $A^{(n)}_{l} = C_l -iS_l$, where $C_l$ and $S_l$ are the (real) coefficients of Fourier cosine and sine transform, respectively. As in the last section, the first-order correction to the mean flow $A^{(1)}_0$ is zero. The nominal locations of UR are now given by
\begin{align}
\label{eq:fullpot3}
r_2 = \left(1 \pm 1/l \right)^{2/3} r_0.
\end{align}
Note that $l$ is now \emph{not} directly associated with any particular $m$. This is because every Fourier mode contributes to the nonlinear driving. From now on, we use $l$ to denote the wavenumber of second-order ultraharmonics modes.

\subsection{The $l$-th harmonics}

The second-order $l$-th harmonics of Equations \eqref{eq:higher3a}-\eqref{eq:higher3c} can be expressed as
\begin{align}
\label{eq:lh_1}
&-i\Omega_l u_l - 2\Omega v_l + c^2 \frac{d}{dr}\sigma_l = R_{r,l}, \\
\label{eq:lh_2}
&-i\Omega_l v_l +2B u_l + \frac{ilc^2}{r}\sigma_l = R_{\varphi,l}, \\
\label{eq:lh_3}
&-i\Omega_l \sigma_l + \frac{du_l}{dr}+u_l\frac{d}{dr}\ln(r\Sigma_0) + \frac{il}{r} v_l = R_{c,l},
\end{align}
where $\Omega_l = l(\Omega_p - \Omega)$ and we drop the superscript $(2)$ for clarity. The driving terms for non-zero $l$ read
{\footnotesize
\begin{align}
\nonumber
R^{(2)}_{r,l} =&-\frac{1}{2}\sum_{m=1}^l \left[ u_{m} \frac{du_{l-m}}{dr} +\frac{i(l-m)v_{m}u_{l-m}}{r} - \frac{v_{m}v_{l-m}}{r} \right] \\
\nonumber
&-\frac{1}{2} \sum_{m=l+1} \left[ u_{m} \frac{du^*_{m-l}}{dr} +\frac{i(m-l)v_{m}u^*_{m-l}}{r} - \frac{v_{m}v^*_{m-l}}{r} \right] \\
&-\frac{1}{2} \sum_{m=1} \left[ u^*_{m} \frac{du_{m+l}}{dr} +\frac{i(m+l)v^*_{m}u_{m+l}}{r} - \frac{v_{m+l}v^*_{m}}{r} \right],
\label{eq:lh_35a}
\end{align}
\begin{align}
\nonumber
R^{(2)}_{\varphi,l} =& -\frac{1}{2}\sum_{m=1}^l\left[ u_{m} \frac{dv_{l-m}}{dr} + \frac{i(l-m)v_{m}v_{l-m}}{r} + \frac{u_{m} v_{l-m}}{r}\right] \\
\nonumber
&-\frac{1}{2} \sum_{m=l+1} \left[u_{m} \frac{dv^*_{m-l}}{dr} + \frac{i(m-l)v_{m} v^*_{m-l}}{r} + \frac{u_{m} v^*_{m-l}}{r}\right]\\
&-\frac{1}{2} \sum_{m=1} \left[u^*_{m} \frac{dv_{m+l}}{dr} + \frac{i(m+l) v_{m+l} v^*_{m}}{r} + \frac{u^*_{m} v_{m+l}}{r}\right],
\label{eq:lh_35b}
\\
\nonumber
R^{(2)}_{c,l} =& - \frac{1}{2r\Sigma_0}\sum_{m=1}^l \frac{d}{d r} (r \Sigma_{m} u_{l-m}) - \frac{il}{2r\Sigma_0}\sum_{m=1}^l \Sigma_{m} v_{l-m} \\
\nonumber
&-\frac{1}{2r\Sigma_0}\sum_{m=l+1} \left[ \frac{d}{d r} (r \Sigma_{m} u^*_{m-l}) - il \Sigma_{m} v^*_{l-m} \right] \\
&-\frac{1}{2r\Sigma_0}\sum_{m=1} \left[ \frac{d}{d r} (r \Sigma^*_{m} u_{m+l}) - il \Sigma^*_{m} v_{m+l} \right].
\label{eq:lh_35c}
\end{align}}
Here we make use of the property $u_{-m} = u^*_m$ for a real variable such that the sum is expressed in terms of positive wavenumber modes only. Although the above expressions for $R^{(2)}_{r,l}$, $R^{(2)}_{\varphi,l}$, and $R^{(2)}_{c,l}$ look complicated, they are simply the $l$-th Fourier component of the right-hand side in Equation \eqref{eq:higher3}. For now, we postpone the discussion of the driving term to Section \ref{sec:force} and derive the equations near the URs where these waves are excited.

By rearranging terms in Equations \eqref{eq:lh_1} and \eqref{eq:lh_2}, we have
\begin{subequations}
\begin{align}
\label{eq:lh_4a}
u_l &= -\frac{i}{D_l}\left[-\Omega_l c^2 \frac{d\sigma_l}{dr} + \frac{2l\Omega c^2}{r}\sigma_l + W_{r,l} \right], \\
\label{eq:lh_4b}
v_l &= \frac{1}{D_l}\left[ 2B c^2 \frac{d\sigma_l}{dr} - \frac{l\Omega_l c^2}{r}\sigma_l + W_{\varphi,l} \right],
\end{align}
\end{subequations}
where
\begin{subequations}
\begin{align}
\label{eq:lh_5a}
D_l &= \kappa^2 - \Omega_l^2, \\
\label{eq:lh_5b}
W_r &= \Omega_l R_{r,l} + 2i\Omega R_{\varphi,l}, \\
\label{eq:lh_5c}
W_\varphi &= -2B R_{r,l} - i\Omega_l R_{\varphi,l}.
\end{align}
\end{subequations}
Note that $D_l(r_2) = 0$ and so $D_l$ is analogous to the resonant denominator $D$ for LRs. Equation \eqref{eq:lh_3} can be written as 
\begin{align}
\label{eq:lh_6}
\Omega_l \sigma_l + \frac{d}{dr}(iu_l) + (iu_l) \frac{d}{dr}\ln(r\Sigma_0) - \frac{l}{r}v_l = iR_{c,l}.
\end{align}
After some algebra, we have the following differential equations \citepalias[c.f.,][]{1992ApJ...389..129A},
\begin{align}
\label{eq:lh_7}
\frac{d^2 \sigma_l}{dr^2}+P(r)\frac{d\sigma_l}{dr}+N(r)\sigma_l - \frac{D_l}{c^2}\sigma_l = R(r),
\end{align}
where
\begin{subequations}
\begin{align}
\label{eq:lh_8a}
P(r) &= \frac{d}{dr}\ln \left(\frac{r\Sigma_0 c^2}{D_l}\right), \\
\label{eq:lh_8b}
N(r) &= \frac{2\Omega}{r(\Omega-\Omega_p)}\frac{d}{dr}\ln\left(\frac{\Sigma_0 \Omega c^2}{D_l}\right) - \frac{l^2}{r^2}, \\
\label{eq:lh_8c}
R(r) &= \frac{1}{\Omega_l c^2}\left[ W_r \frac{d}{dr}\ln\left(\frac{r\Sigma_0 W_r}{D_l}\right)-\frac{l}{r}W_\varphi -iD_l R_{c,l} \right].
\end{align}
\end{subequations}
Equation \eqref{eq:lh_7} is analogous to the governing equation in \citetalias{1992ApJ...389..129A} except that the driving terms and EOS are different.

\subsection{Ultraharmonics Resonances}

Near the URs, a dimensionless coordinate can be defined by
\begin{align}
\label{eq:ur_2}
x = \frac{r-r_2}{r_2},
\end{align}
where $r_2$ is defined in Equation \eqref{eq:fullpot3}. Similar to the case of LRs, we expand $D_l$ in Taylor series, that is
\begin{align}
\label{eq:ur_3}
D_l = \mathscr{D}_l x,
\end{align}
where $\mathscr{D}_l = (rd D_l / dr)|_{r_2} = -3(1\pm l)\Omega_l^2$ for a Keplerian disk. Thus, we have
\begin{align}
\label{eq:ur_4}
r\frac{d}{dr}\ln D_l \simeq \frac{1}{x} + C_0,
\end{align}
where $C_0=\pm (3l/2)-4$.
Equations \eqref{eq:lh_8a}-\eqref{eq:lh_8c} can be expanded in a Laurent series in $x$. Similar to the equations near LRs \citep{1979ApJ...233..857G}, we define a length scale
\begin{align}
\label{eq:ur_5}
\Lambda = \mp \left(\frac{c^2}{r_2^2 |\mathscr{D}_l|}\right)^{1/3},
\end{align}
which becomes apparent when solving for an analytical solution. By further defining a dimensionless variable $X=x/\Lambda$, Equation \eqref{eq:lh_7} can be written as (\citetalias{1992ApJ...389..129A}),
\begin{align}
\nonumber
&\left[\frac{d^2}{dX^2}-\left(\frac{1}{X}-\varepsilon\right)\frac{d}{dX}-\gamma\left(\frac{1}{X}-\varepsilon+\frac{5\Lambda}{2}\right)-\frac{\gamma^2}{4}\right]\sigma_l \\
\label{eq:ur_6}
&- X\sigma_l = S(X),
\end{align}
where
\begin{subequations}
\begin{align}
\label{eq:ur_7a}
\varepsilon &= \Lambda \left( 1 + \frac{d\ln\Sigma_0}{d\ln r} + \frac{d\ln c^2}{d\ln r} - C_0 \right),\\
\label{eq:ur_7b}
\gamma &= \frac{2l\Omega|\Lambda|}{\kappa},\\
\label{eq:ur_7d}
S(X) &= r_2^2 \Lambda^2 R.
\end{align}
\end{subequations}
Coefficients in Equation \eqref{eq:ur_6} are evaluated at $r=r_2$. For a particular cold disk model with $c^2=c_0^2/r$ and aspect ratio $h_0=c_0/r_0\Omega_p = 0.05$, $|\Lambda| \simeq (l \pm 1)^{-1/3} (h_0^2/3)^{1/3} \lesssim 0.07$ for $l>1$. As noted in \citetalias{1992ApJ...389..129A}, $\epsilon$ and $\gamma$ may be neglected, which affect only the details of the waveform. Assuming $\gamma=\varepsilon=0$, we have
\begin{align}
\label{eq:ur_8}
\frac{d^2\sigma_l}{dX^2}-\frac{1}{X}\frac{d\sigma_l}{dX}- X\sigma_l = S_0(X),
\end{align}
where $S_0(X)$ is the approximation to $S(X)$ and we suppress the label of $l$ for clarity. Equation \eqref{eq:ur_8} is the governing equation of the ultraharmonics waves of a particular $l$. Therefore, it is important to get an expression of $S_0(X)$. In the following section, we extend the analysis to study a general form of nonlinear mode-coupling.

\section{Driving due to Nonlinear Mode-coupling}
\label{sec:force}

In this section, we investigate the driving term $S_0(X)$ in Equation \eqref{eq:ur_8}, which is a linear combination of $R^{(2)}_{r,l}$, $R^{(2)}_{\varphi,l}$, and $R^{(2)}_{c,l}$ in Equations \eqref{eq:lh_35a}-\eqref{eq:lh_35c}. We begin by discussing their general properties. First, they are generally non-zero for each $l$. Therefore, the second-order solution (e.g., $u^{(2)}$) is a superposition of all $l$-modes. Second, the driving terms are the $l$-th Fourier component of the quadratic terms. In Section \ref{sec:source} below, we discuss the importance of such driving terms, which allow preferential excitation to low-$l$ harmonics response. 

It can be shown that radial wavenumber of the driving term leads to a phase shift in the second-order waves \citepalias{1992ApJ...389..129A}. In the following, we investigate the radial wavenumber and the phase of the driving term. To proceed, we consider the following expression that represents summations in the driving term $R$,
\begin{align}
\nonumber
&R_l = \frac{1}{2\pi}\int^\infty_{-\infty} f(\varphi)g(\varphi) e^{-il\varphi}d\varphi \\
\label{eq:force0}
&= \frac{1}{2}\sum^l_{m=1} f_m g^*_{l-m} + \frac{1}{2}\sum^\infty_{m=l+1} f_m g^*_{m-l} + \frac{1}{2}\sum^\infty_{m=1} f^*_m g_{m+l},
\end{align}
where $f_m$ and $g_m$ are the Fourier components of functions $f$ and $g$. The phase of $R_l$ is the same as that of the true driving term under WKB approximation. Since the driving term is to be evaluated at the URs, $r_2=(1\pm 1/l)^{2/3}r_0$, a $m$-th linear solution contributes to the forcing with its non-wave and wave-like responses for $l \geq m$ and $l<m$, respectively. From Equation \eqref{eq:force0}, we identify there are four cases of interactions between linear solutions. These cases can be found by comparing the resonance locations among wavenumbers $l$, $m$ and $|l\pm m|$. 

\begin{figure*}
\includegraphics[width=\textwidth]{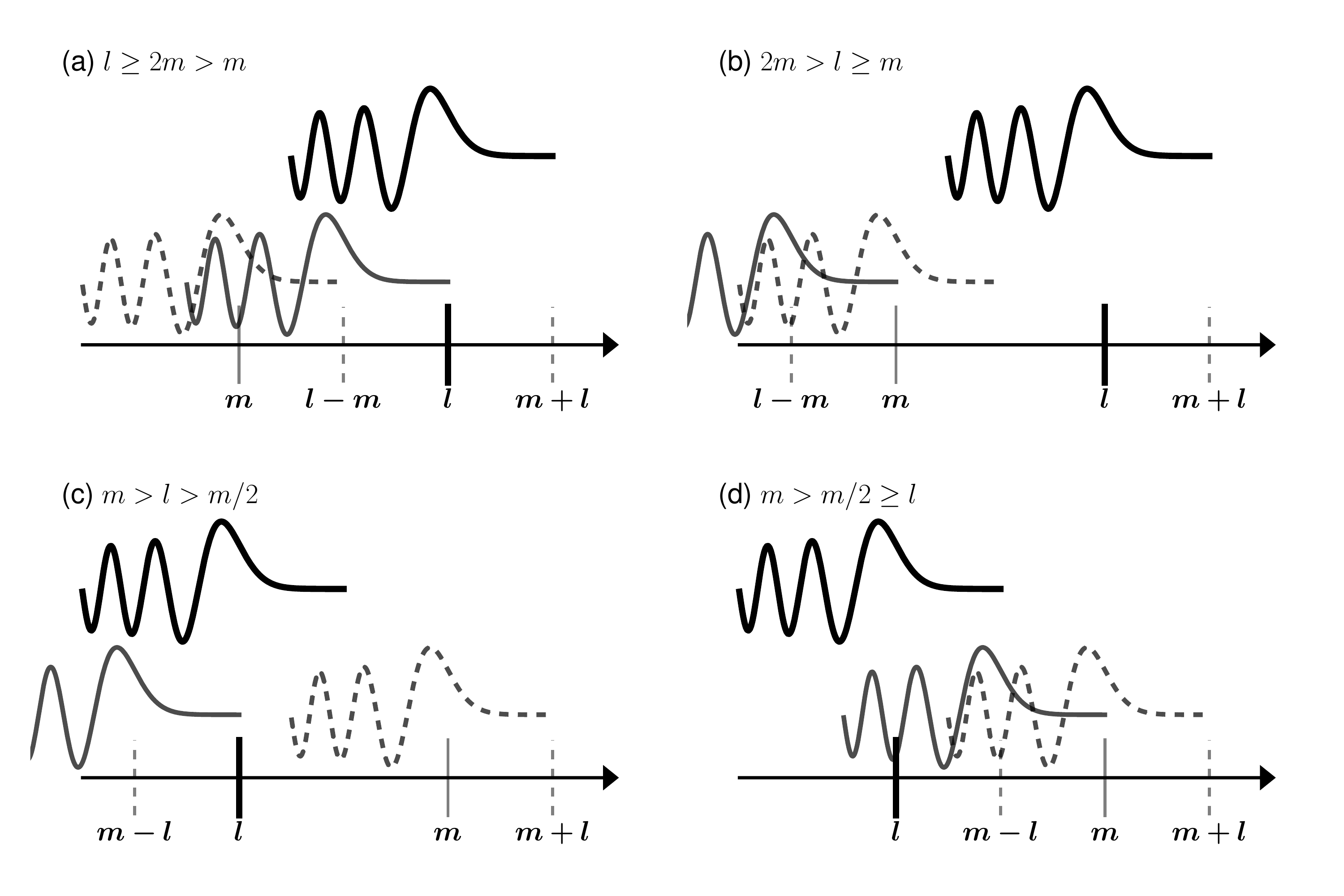}
\caption{An illustration showing four types of mode-couping between linear waves (see Section \ref{sec:force}). The linear waves excited at LRs (grey and dashed lines) can interact at the $l$-th UR (i.e., $r_2$) to excite ultraharmonics wave (solid black line). The axis indicates the direction towards the planet (but the distance is not in scale). The tick marks represent the $l$-th UR (thick) and other LRs. The four panels are (a) $r_2 > r_L^{l-m} >r_L^{m}$, (b) $r_2 > r_L^{m} >r_L^{l-m}$, (c) $r_L^{m-l} > r_L^{m} > r_2$, and (d) $r_L^{m} > r_L^{m-l} > r_2$, where $r_L^m$ is the $m$-th LR. For clarity, the $(m+l)$-th mode is not shown, which always contribute with its wave-like response at $r_2$.}
\label{fig:wave1}
\end{figure*}

To understand better the mode-coupling, an illustration demonstrating four types of interaction between linear waves is shown in Figure \ref{fig:wave1}. According to Equation \eqref{eq:force0}, each $m$-th linear mode interacts with both $(m+l)$-th mode and $|m-l|$-th mode. When $m \leq l$, the $m$-th and $(l-m)$-th modes interact at $r_2$ with their non-wave response. In Figure \ref{fig:wave1}, panels (a) and (b) represent the cases for $l-m > m $ and $l-m \leq m $, respectively. When $m>l$, the $m$-th mode contributes with its wave-like response at $r_2$, whereas the contribution of $(m-l)$-th mode depends on its relative location to $r_2$. The case (c) and (d) correspond to $m-l > l $ and $m-l \leq l $, respectively. In all four cases, the $(l+m)$-th mode contribute with its wave-like response at $r_2$ since $l+m > l$ for all $m$.

In general, the interactions between $m$-th and $|m\pm l|$-th modes can be characterized into three types, namely, ``wave-wave", ``wave-non-wave", and ``non-wave-non-wave". In the following, we proceed to determine the resultant phase of these terms. Consider a quadratic term which consists of two linear waves, e.g.,
\begin{align}
\label{eq:force1}
u_{m+l} u^*_m = U_{m+l}U_m^* \exp i\Phi_{m,l}(r) ,
\end{align}
where $\Phi_{m,l}$ is the total phase of the product and the constant phase in Equation \eqref{eq:linear1} is canceled in $U_{m+l}U_m^*$. In this case, we require $m>l$ for both $u_{m+l}$ and $u^*_m$ to be wave-like at $r_2$ (i.e., case (c)). The relevant phase is given by
\begin{align}
\label{eq:force2}
\Phi_{m,l}(r) = \int^r_{r^{m+l}_L} k_{m+l}(s) ds - \int^r_{r^{m}_L} k_m(s) ds,
\end{align}
where $r^m_L$ is the $m$-th LRs. This term appears as the two waves are launched from different LRs and interacting at a particular UR. Expanding it in a Taylor's series about $r_2$, we have
\begin{align}
\label{eq:force3}
\Phi_{m,l}(r) &= \Phi_{m,l}(r_2) + q^{(l)}_m X,
\end{align}
where $q^{(l)}_m = (k_{m+l}-k_m)r_2\Lambda$ is the effective wavenumber of the driving terms evaluated at $r_2$.

For ``wave-non-wave" interaction, only the wave-like response contributes to the phase and effective wavenumber. Consider a quadratic term ($u_{m+l} u^*_m$) for $l>m$, we have
\begin{align*}
\Phi_{m,l}(r_2) = \int^{r_2}_{r^{m+l}_L} k_{m+l}(s) ds \quad \text{and} \quad q^{(l)}_m = k_{m+l}r_2\Lambda,
\end{align*}
where only the wave-like $u_{m+l}$ is responsible. Finally, the ``non-wave-non-wave" interaction result in zero phase and zero effective wavenumber. Next we use dispersion relation in Equation \eqref{eq:linear2} to obtain the expressions for $\Phi_{m,l}(r_2)$ and $q^{(l)}_m$.

\subsection{Estimation of the Phase and Effective Wavenumber}
\label{sec:source}

To facilitate the derivation, we adopt a particular disk model with $c^2 = c_0^2/r$ (i.e., $\beta=1$). Using Equation \eqref{eq:linear2} and assuming a Keplerian disk, the wavenumber of a $m$-th linear waves at $l$-th UR (i.e., $m > l$) is given by
\begin{align}
\label{eq:force4}
k_m(r_2) = \frac{1}{r_2 h_0}\left(\frac{m^2}{l^2}-1\right)^{1/2},
\end{align}
where $h_0 = c_0/r_0\Omega_p$ is the constant aspect ratio in this model. Here we first consider the ``wave-wave" interaction. The effective wavenumber reads
\begin{align}
\nonumber
&q^{(l)}_m = (k_{m+l}-k_m)r_2\Lambda \\
\label{eq:force5}
&= \frac{\Lambda}{h_0}\left[ \left(\frac{(m+l)^2}{l^2}-1\right)^{1/2} - \left(\frac{m^2}{l^2}-1\right)^{1/2}\right],
\end{align}
where $\Lambda$ is given by
\begin{align}
\label{eq:force6}
\Lambda = \mp \left[\frac{h_0^2}{3|1\pm l|}\right]^{1/3}.
\end{align}

To get an estimate of the source term and its phase $\Phi_{m,l}(r_2)$, we further approximate the dispersion far away from the $m$-th LR as
\begin{align}
\label{eq:source1}
m^2 (\Omega_p - \Omega)^2 \simeq k_m^2 c^2,
\end{align}
where $\kappa^2 \simeq \Omega^2$ is dropped as it becomes very small compared to the other term. In particular, this approximation is valid when $m \gg 1$. In fact, the ``wave-wave" interaction (Equation \eqref{eq:force2}) only appears when $m>l$ (i.e., cases (c) and (d) in Figure \ref{fig:wave1}). Since the torque contribution is mainly due to large $m$ from the linear theory, this is a relevant approximation for the linear waves for mode-coupling. In any case, we estimate the wavenumber \citepalias{2002MNRAS.330..950O} as,
\begin{align}
\label{eq:source2}
k_m &\simeq \pm \frac{m}{c} (\Omega_p - \Omega) = \pm \frac{m}{h_0} \left(r^{1/2} - r^{-1}\right),
\end{align}
where the plus (minus) sign correspond to the outer (inner) resonance (i.e., $k>0$). The inner and outer LRs are now approximated to be $r_L \simeq 1$ to this order. Thus, the phase and the effective wavenumber of the nonlinear driving are
\begin{align}
\label{eq:source3}
\Phi_{m,l}(r_2) &\simeq \pm \frac{l}{h_0}\left(\frac{2}{3}r_2^{3/2} - \ln r_2 \right),
\end{align}
and,
\begin{align}
\label{eq:source4}
q^{(l)}_m &\simeq  \pm \frac{l}{h_0} \left(r_2^{1/2} - r_2^{-1}\right)r_2\Lambda = \frac{\Lambda}{h_0},
\end{align}
respectively, which are both independent of $m$ in the large-$m$ limit.

Similarly, $\Phi_{m,l}(r_2)$ and $q^{(l)}_m$ can be obtained for the ``wave-non-wave" interaction. For $m \leq l$ (i.e., cases (a) and (b)), the relevant wavenumber is $k_{l+m}$ (only $u_{m+l}$ is wave-like). For $m>l$ (case (c) only), the relevant wavenumber is $k_{m}$. Using Equation \eqref{eq:force4}, we have 
\begin{align}
\label{eq:source5}
q^{(l)}_m = k_{m}(r_2)r_2\Lambda = \frac{\Lambda}{h_0}\left(\frac{m^2}{l^2}-1\right)^{1/2},
\end{align}
where $m$-th mode is wave-like at $r_2$. The phase $\Phi_{m,l}$ for the ``wave-non-wave" interaction can be obtained without the approximation in Equation \eqref{eq:source1} and by making use of an analytical expression presented in \citetalias{2002MNRAS.330..950O}. Since the expression is not illuminating, we proceed without giving an explicit expression. 

As noted above, the phase and wavenumber for ``non-wave-non-wave" interactions are zero. As a result, their contribution to $S_0(X)$ takes a form similar to the planet's potential that is responsible for the first-order modes, which can be solved in terms of Airy functions \citep[e.g.,][]{1987Icar...69..157M}.

In summary, for a pair of $m$ and $l$, the mode-coupling takes place between $m$-th and $|m\pm l|$-th modes. Here we denote $q^{(l)+}_{m}$ and $q^{(l)-}_{m}$ for the effective wavenumber due to the interactions with $(m+l)$-th and $|m-l|$-th modes, respectively. Thus, we have
\begin{align}
\label{eq:source7a}
q^{(l)+}_{m}= 
\begin{cases} 
\Lambda/h_0\left(\frac{(m+l)^2}{l^2}-1\right)^{1/2} & \text{if } m \leq l \\
\Lambda/h_0 & \text{if } m > l \\
\end{cases},
\end{align}
and
\begin{align}
\label{eq:source7b}
q^{(l)-}_{m}= 
\begin{cases} 
0 & \text{if } m \leq l \\
\Lambda/h_0\left(\frac{m^2}{l^2}-1\right)^{1/2} & \text{if } m/2 < l < m \\
\Lambda/h_0 & \text{if } l \leq m/2 \\
\end{cases},
\end{align}
where we assume $m \gg 1$. Similarly, the phase $\Phi^\pm_{m,l}(r_2)$ can be obtained by considering the corresponding type of interaction discussed above. 

To conclude this section, we note that the resultant driving term in the form of Equation \eqref{eq:force0} can be understood as combination of two types of interactions, namely, the interactions between $m$-th and $|m\pm l|$-th modes. Therefore, the driving term can be further simplified and allow us to solve Equation \eqref{eq:ur_8} analytically. We describe the method of solution in Appendix \ref{sec:methodofsolution}. Here, we quote the result of a WKB expression of the second-order wave ($\sigma_l$) (see Equation \eqref{eq:solve11}). The phase of $\sigma_l$ near the URs can be expressed as
\begin{align}
\label{eq:source6}
\Phi_{m,l}(X) &= \Phi_{m,l}(r_2) - \frac{(q^{(l)}_m)^3}{3} - \frac{2}{3}(-X)^{3/2}  + l\varphi + \frac{\pi}{4},
\end{align}
which corresponds to a trailing wave that propagates towards $X \rightarrow -\infty$. Finally, in addition to the phase-shift $(q^{(l)}_m)^3/3$ due to wave-like forcing terms discussed in \citetalias{1992ApJ...389..129A}, we note that there is a non-trivial phase $\Phi_{m,l}(r_2)$ present in the expression.

\subsection{Disk Response to the Nonlinear Driving}
Without directly integrating Equation \eqref{eq:lh_7}, some qualitative results can still be obtained from previous derivations. By solving the simplified governing equation (Equation \eqref{eq:ur_8}), the excitation of the individual ultraharmonics wave can be studied. In particular, the driving term $S_0$ is a summation of products of linear modes. It can be expressed in a convolution form, (e.g., sum of $u_{m+l}u^*_m$ among all $m$). Indeed, this is an important property: the terms with small wavenumber difference (i.e., $l$) contributes more to the sum. This property can be easily understood by making analogy to a convolution integral for continuous functions, where the contribution is the largest when there is substantial overlapping between the functions. Therefore, the driving term $R$ generally favors low-$l$ ultraharmonics.

On the other hand, unlike the planet's potential which depends on the distance to it, the nonlinear driving terms in Equations \eqref{eq:lh_35a}-\eqref{eq:lh_35c} depend explicitly on $r$, which is the distance to the central star. Therefore, the forcing is expected to be stronger for the inner-most URs (i.e., $l=2,3$). Although the driving term for $l=1$ may be large, the inner resonance is located at the origin. Therefore, even such mode is excited, it is a non-wave (i.e., non-spiral) response in the inner disk. We leave this special case for future study.

The only driving term that appears for a single $m$ potential in Section \ref{sec:Section3}, i.e., $l=2m$, turns out to be not important. It corresponds to case (a) discussed in Section \ref{sec:force}. Such term only contributes once in each summation series over $m$ in $R$ for a particular $l$. Since the phase of other modes (due to wave-like driving) under the large $m$ approximation does not depends on $m$ (see Equation  \eqref{eq:source4}), they may be easy to have constructive interference.

\subsubsection{Difference between Inner Disk and Outer Disk}

An analogous statement regarding the linear wave amplitude can also be constructed for the ultraharmonics waves. Using Equation \eqref{eq:lh_7},\eqref{eq:lh_8a},\eqref{eq:source1}, and WKB approximation for free waves, we have $S_l \propto k_l^{1/2}$, where $S_l$ is the wave amplitude and $k_l$ is the wavenumber of pressure waves (by replacing $m$ by $l$ in the dispersion relation). Therefore, similar to the results from linear theories, the amplitude of ultraharmonics waves generally increases along propagation, where the wave amplitude in the inner disk is generally larger as the wavenumber $k$ increases more rapidly.

As the indirect potential due to the offset of the center of mass is excluded in this study, the $m=1$ mode for the linear solution is absent \citep[c.f.,][]{1990ApJ...358..495S}. However, since the nonlinear driving contains $l=1$ mode, one may still expect such ultraharmonics wave to be launched at the outer UR, $r_2 = 2^{2/3} r_0$.

\section{Numerical Results}
\label{sec:Section6}

In this section, we examine the mechanism of generating multiple spiral arms by ultraharmonics waves. The gas response to an external potential is studied by performing numerical simulations and using a 2D finite-volume hydrodynamics code in cylindrical coordinates, which is derived from a higher-order Godunov code \textit{Antares} \citep{2005JKAS...38..197Y}. An independent version has been extended to study circum-planetary disk problems using nested grid refinement \citep{2014ApJ...790...32W}. A basic serial version is used in this work. Since the numerical simulations of planet-disk interaction are fairly standard and can be easily performed using publicly available codes, interested readers may refer to Appendix \ref{sec:NumericalCode} for implementation details. Our simulation parameters are based on the work by \citet{2006MNRAS.370..529D}, which compared several hydrodynamics codes. In our simulation, an initial condition of a Keplerian disk is used, which is the solution to Equation \ref{eq:eq0}. The radial profiles for surface density and sound speed are given by
\begin{align}
\Sigma_0 &= \Sigma_{00} (r/r_0)^{-1}, \quad \text{and} \quad c = c_0 (r/r_0)^{-1/2},
\end{align}
respectively, where $\Sigma_{00} = 1$ is the surface density and $c_0=0.05$ is the local sound speed at $r=r_0=1$. We set the inner and outer radius to be 0.4 and 2.5, respectively. Non-reflective boundary condition is used at the radial boundaries. The simulations are performed with $N_r \times N_\varphi = 256 \times 768$ grid cells. Logarithmic grid spacing is used for the radial direction. 

In the following, we first study the case of a single $m$-th component of $\Phi_p$ (i.e., $\phi_m$). Then we study multiple Fourier components and eventually the full potential of a planet. We slowly increase the strength of the potential in a few orbits. For simple implementation, we compute $\phi_m(r)$ by Fast Fourier Transform (FFT) instead of using Laplace's coefficients. A smoothing length $r_s \sim 0.6 H_0$ is used in the simulations. For easier comparison to the linear theory of individual Fourier component, we use the mass ratio to fix the strength of $\phi_m$. This implies that $\phi_m$ is different for different $m$ at the resonances. 

\subsection{Single-Mode Potential}
\label{sec:singlemode}

\begin{figure}
\centering
\includegraphics[width=0.5\textwidth]{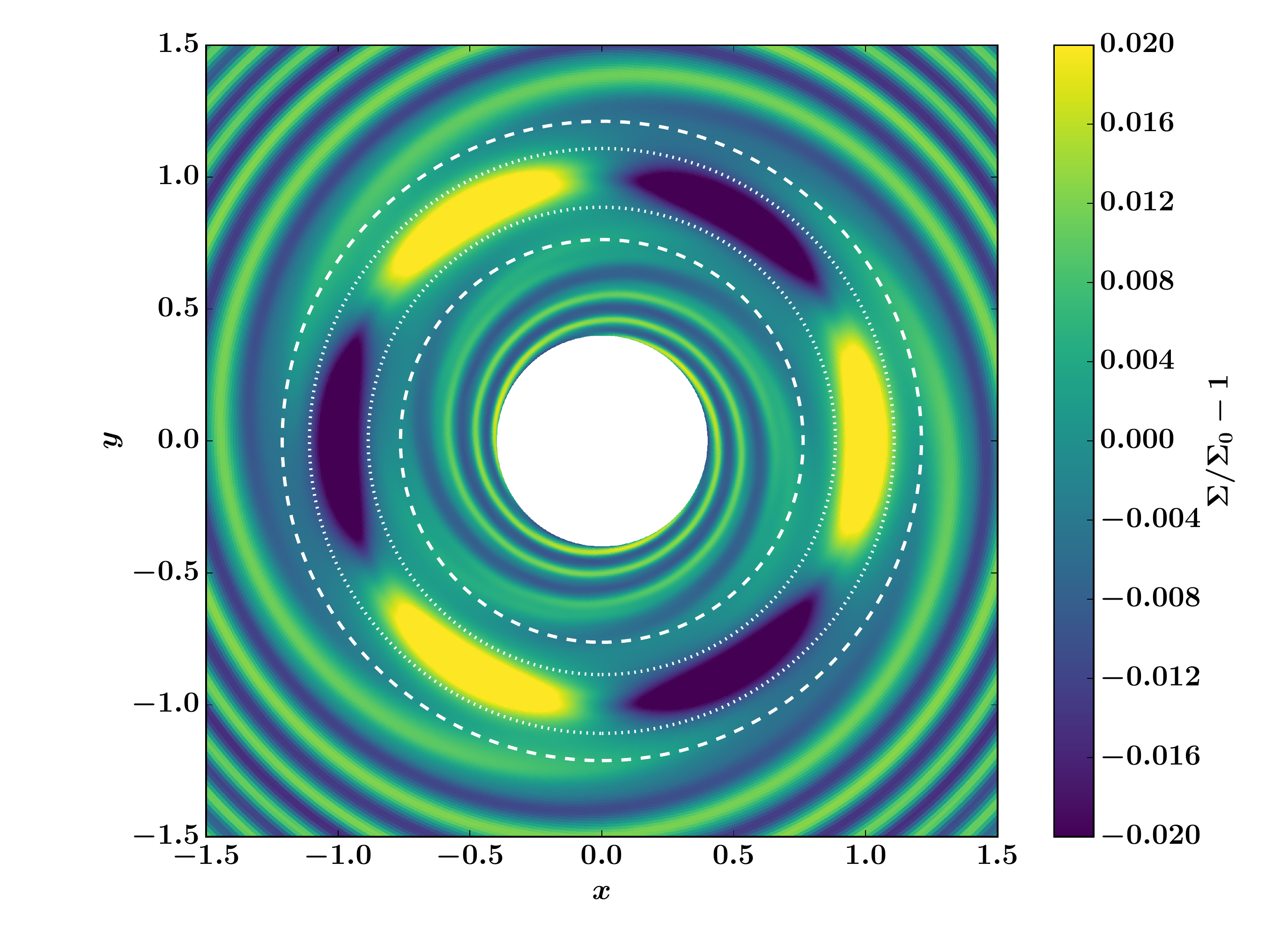}
\caption{Surface density perturbation ($\Sigma/\Sigma_0-1$) for the gas response to a $m=3$ potential. The dashed and dotted lines indicate the locations of the LR ($r_L$) and second-order UR ($r_2$), respectively. The upper and lower limits of color scale are fixed to $\pm 0.02$ to enhance the contrast of the spiral waves.}
\label{fig:densitym3}
\end{figure}
A single mode $\phi_m$ is used as a static external potential in the rotating frame. The linear solution for this case can also be obtained by integrating the ordinary differential equations directly \citep{1993Icar..102..150K,2002ApJ...565.1257T,2003MNRAS.341.1157T}. However, the use of a hydrodynamics code allows us pick up the nonlinear modes, as we show below.

We begin with a $m=3$ component. The result after 8 orbits is shown in Figure \ref{fig:densitym3}. The three-armed spiral structures are excited in both inner and outer disk. The result agrees very well with the linear calculation. In particular, spiral density waves are excited at the LRs with an expected phase-shift of $-\pi/4$ for surface density. The gas response between two LRs is the non-wave part of the solution (see Equation \eqref{eq:linear4}), which is responsible to the excitation of ultraharmonics waves in this case. In the first three rows of Figure \ref{fig5}, we present the Fourier analysis for the gas response for $q=10^{-5}$, $10^{-4}$, and $10^{-3}$. The Fourier amplitude of surface density perturbations $\Sigma/\Sigma_0-1$ at $r=0.55r_0$ are also shown. The ultraharmonics ($m=3n$) are excited, which is expected by the mode-coupling mechanism described in Section \ref{sec:Section3}. All other modes are at $\lesssim 10^{-14}$ level. We note that the first few components agree with a power-law relation, where the $n$-th order amplitude goes roughly $\propto \epsilon^n$. For very high-order modes (i.e., $n \geq 5$ for $q \leq 10^{-4}$), the amplitudes depart from the power-law. It is because the actual amplitudes depend on how far the ultraharmonics waves propagate and the possible higher-order corrections. We note that the resolution may be inadequate for the high-$m$ modes, in which the radial wavelength is small. In any case, the amplitudes of these cases scale roughly with the mass ratio $q$. 

\subsection{Multiple-Mode Potential}
\label{sec:multimode}

\begin{figure}[!b]
\centering
\includegraphics[width=0.5\textwidth]{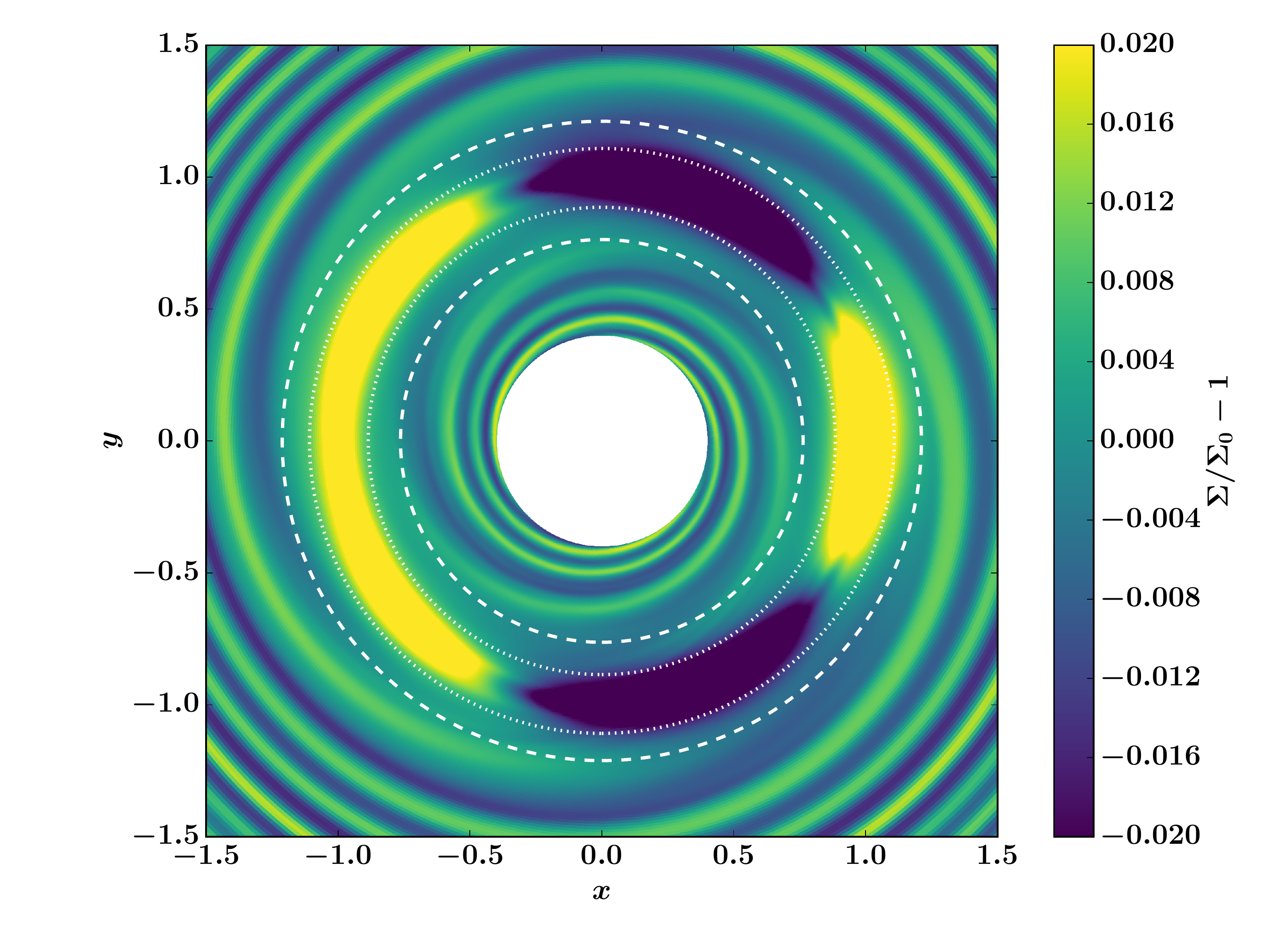}
\caption{Same as Figure \ref{fig:densitym3}, but for $m=2$ and $m=3$ components.}
\label{fig4}
\end{figure}

\begin{figure*}
\centering
\includegraphics[width=\textwidth]{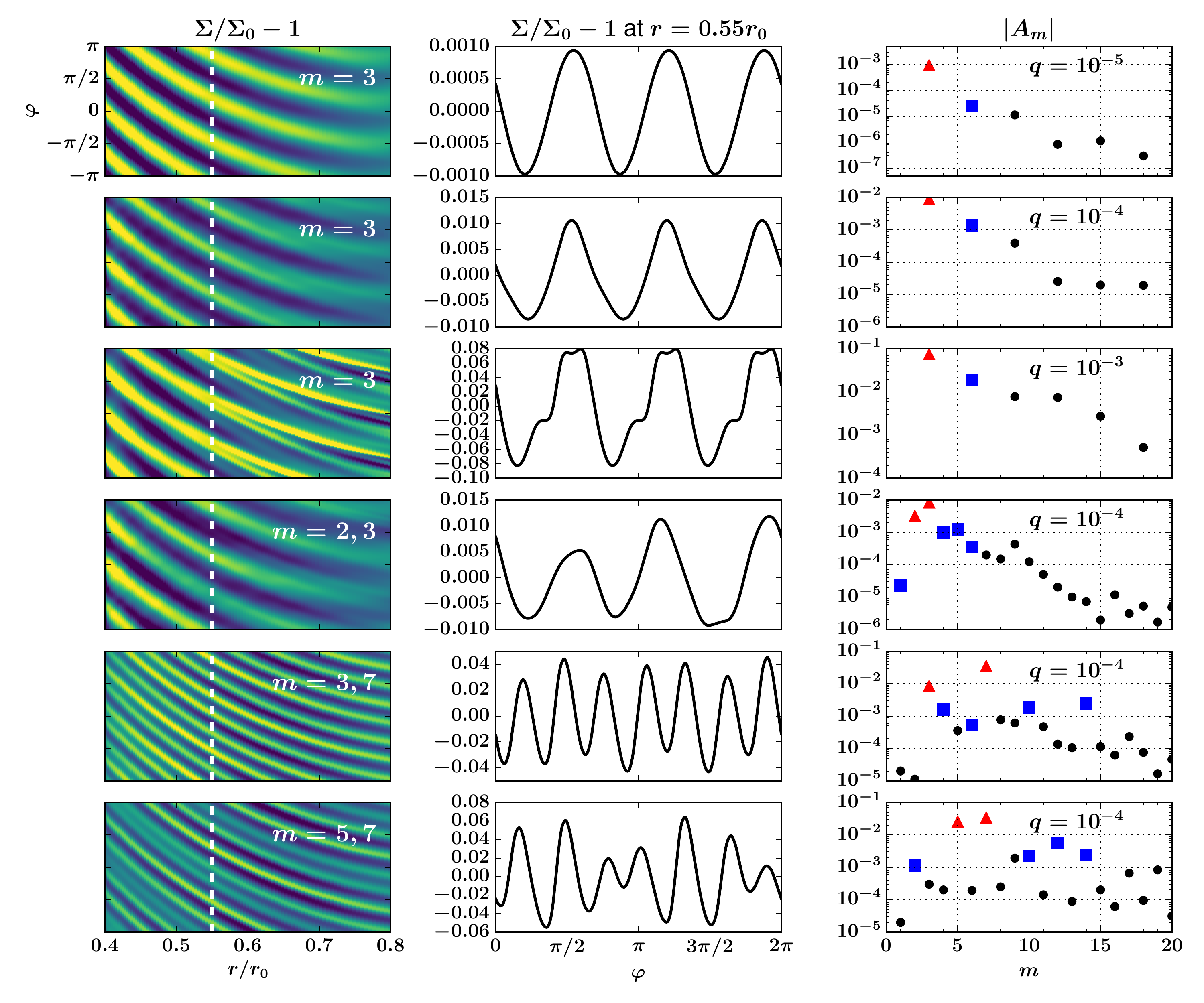}
\caption{Gas response due to single and multiple Fourier components of a planet's potential. The mass ratio from top to bottom rows are $q=10^{-5}$, $10^{-4}$, $10^{-3}$, $10^{-4}$, and $10^{-4}$. The first 3 rows are the cases for $m=3$. The fourth, fifth, and sixth rows are for $m=2,3$, $m=3,7$, and $m=5,7$, respectively. The left column shows the surface density perturbation in linear color scale (not to scale w.r.t. each row). The white dashed line marks the location of $r=0.55r_0$. The middle column shows the azimuthal profile of surface density perturbation at $r=0.55r_0$. The right column shows the spectrum of Fourier amplitudes $|A_m|$ (up to $m=20$ are shown). The red triangles and blue square indicates the first and second-order modes, respectively.}
\label{fig5}
\end{figure*}

When multiple Fourier components are considered, there is a huge variety of modes present in the gas response. In Section \ref{sec:Section4}, we find that the forcing of $n$-th order modes consists of $m\pm m'$-th harmonics, where $m$ and $m'$ are any combinations of the Fourier components of the $n'$-th and $(n-n')$-th order modes. To demonstrate this effect, we study a combination two components of a planet's potential (Figure \ref{fig4}). Since we fix the mass ratio, the linear amplitude of larger $m$ is stronger. We choose a radius $r=0.55r_0$ and study the Fourier components of the azimuthal profile of the surface density perturbation (right column). The result is shown in last three rows Figure \ref{fig5}. Here we study the cases for three pairs, namely, $m=(2,3)$, $(3,7)$, and $(5,7)$. As expected, the two most dominant modes in the inner disk are the first-order modes (red squares). Due to mode-coupling, the expected second-order modes (blue squares) are strongly excited as well, which are among the highest amplitudes in the simulation. The second-order modes for $m=(2,3)$, $(3,7)$, and $(5,7)$ are $l=(1,4,5,6)$, $(4,6,10,14)$, and $(2,10,12,14)$, respectively. We note that even the higher-order $l=1$ is expected from coupling between $m=2,3$ (see fourth rows of Figure \ref{fig5}), it is only weakly excited.

\subsection{Full Planet's Potential}
\label{sec:fullplanet}

\begin{figure*}
\includegraphics[width=\textwidth]{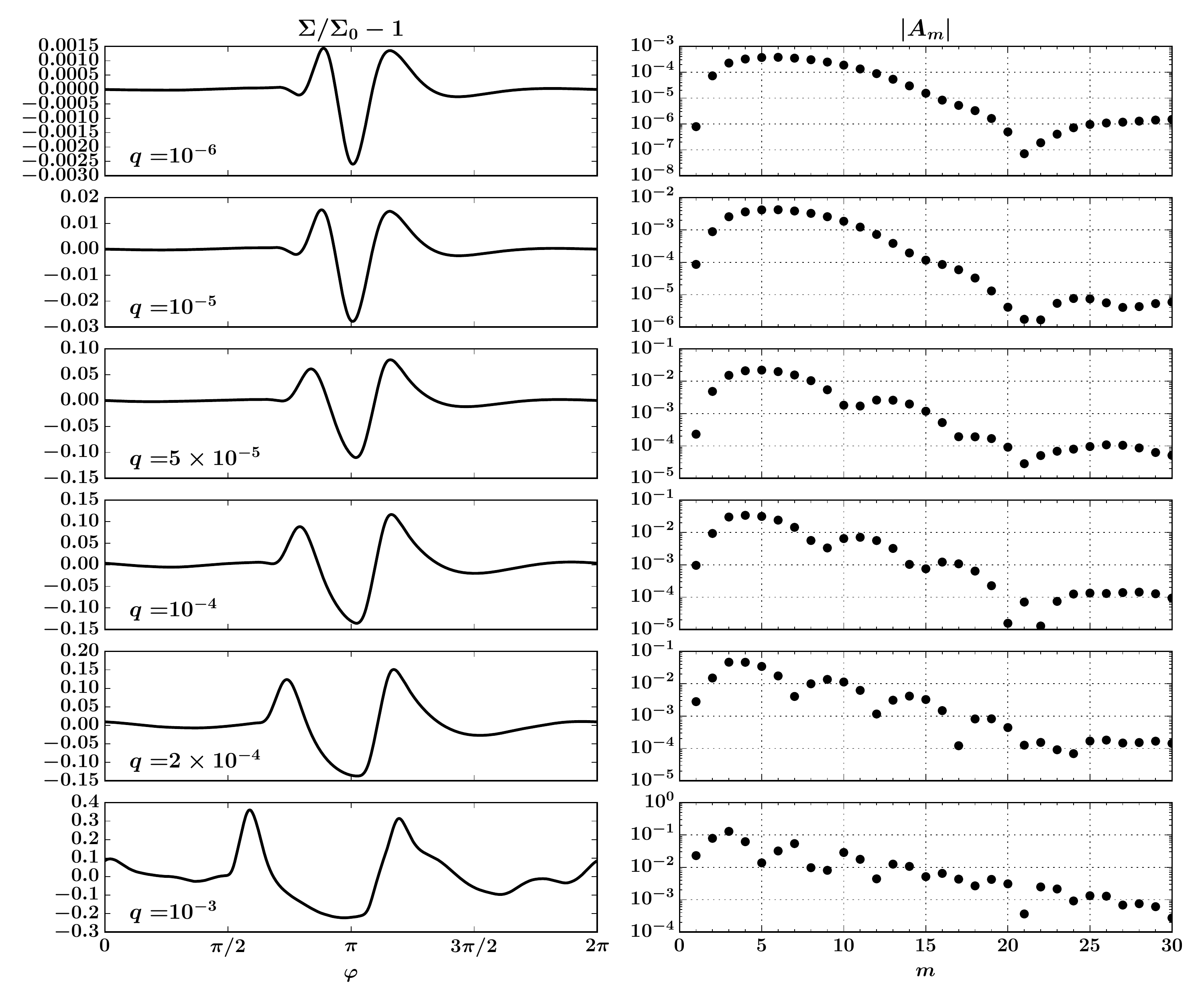}
\caption{The gas response of different mass ratios $q$. The mass ratios from the top to bottom rows are $q=10^{-6}$, $10^{-5}$, $5\times 10^{-5}$, $10^{-4}$, $2\times10^{-4}$, and $10^{-3}$. The left column shows the surface density perturbation ($\Sigma/\Sigma_0-1$) in the azimuthal direction at a particular radius $r=0.6r_0$. The right column shows the Fourier amplitudes.}
\label{fig:qcases}
\end{figure*}

We now consider the full planet's potential. Same set of parameters are used, including the smoothing length. In Figure \ref{fig:qcases}, we show the gas response for various $q$ at $r=0.6r_0$ with averaging over 5 neighboring grid cells. The data are selected at different time for each $q$ about 15-50 orbits, generally with more orbits for higher $q$ in order to reach a quasi-steady state. On the right column of the same figure, we show the amplitude of Fourier components in log-scale up to $m=30$. We note that, in the linear regime $q \leq 10^{-4}$, the gas response of each case generally scales linearly with $q$. The Fourier amplitudes have a flat peak around $m \simeq 5$. This value is different from what would be expected from the wave torque calculation (i.e., $m\simeq 1/(2h_0)=10$) \citep{1980ApJ...241..425G,1993ApJ...419..155A} because we are measuring the amplitude at a particular non-resonant radius. Amplitude change due to propagation should be taken into account in order to make a quantitative comparison (c.f., Equation \eqref{eq:linear3}). As we increase the planet's mass, the separation between two spiral waves (i.e., two peaks on the left column) increases. Our result generally agrees with \citetalias{2015ApJ...815L..21F} for similar scale-height and planet's mass. On the other hand, the density perturbations are generally weaker compared to the two-dimensional cases in \citet{2015ApJ...813...88Z} because a lower mass planet and a hotter disk are used.

Another feature in Figure \ref{fig:qcases} is that the Fourier spectrum becomes more straight (i.e., a power-law) when mass increases. We speculate this change is due to nonlinear mode-coupling described in Section \ref{sec:force} which shifts the spectrum to lower $m$. The details of how the spectrum transits from one that is determined by the wave torque in linear theory to one that is heavily modified by mode-coupling is left for further study. We note that this comparison will require detail calculation of the wave amplitude and phase at URs, which are ignored in this paper. In any case, such a calculation will help understand quantitatively the power-law relation between separation of spirals and mass ratio found by \citetalias{2015ApJ...815L..21F}.

\subsection{Mode-coupling for Linear Waves}
\label{sec:modecouplelinear}

\begin{figure}
\includegraphics[width=0.5\textwidth]{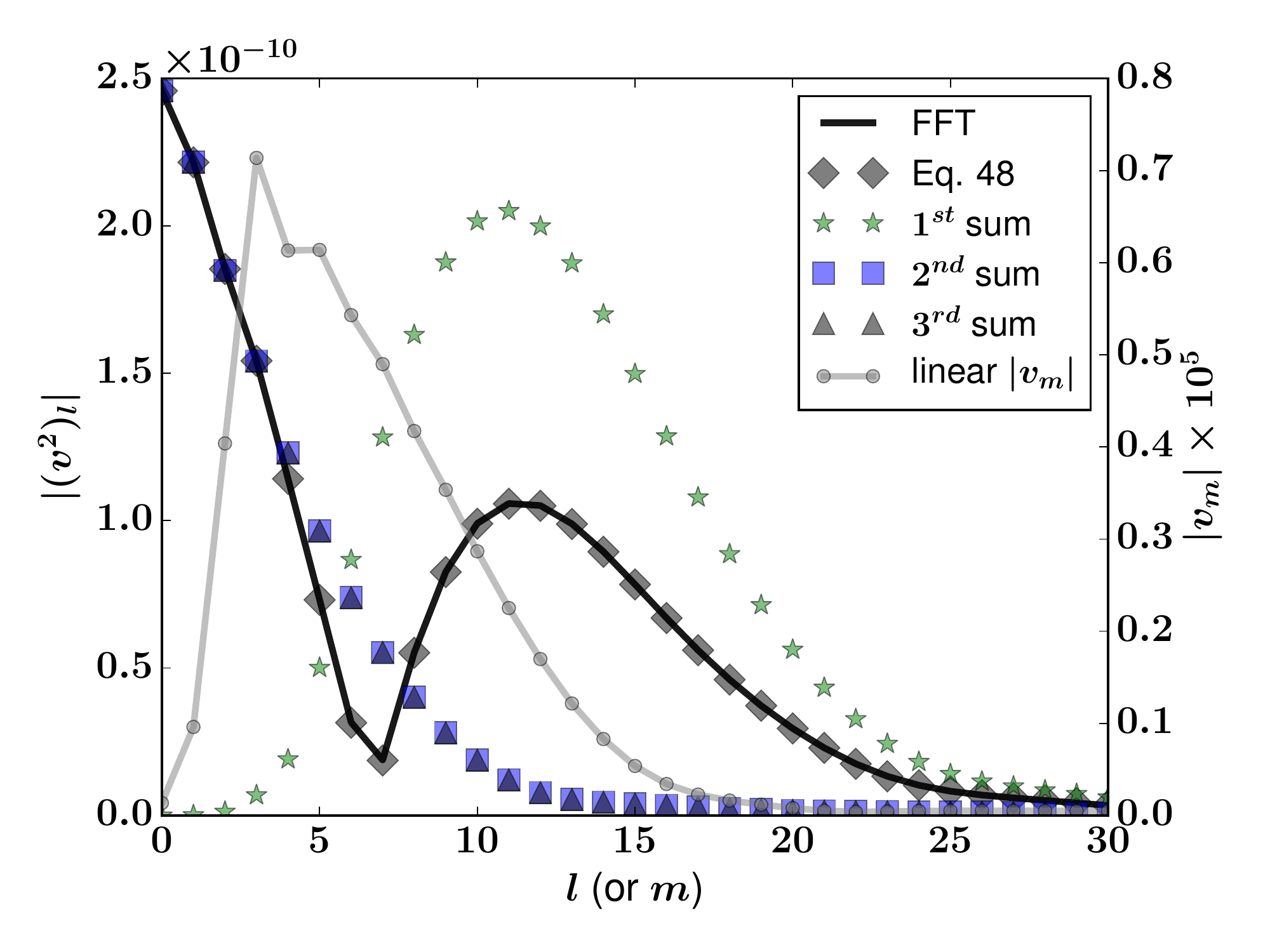}
\caption{A plot of Fourier amplitudes of $v^2$ obtained by FFT (black line) and by summing over linear modes (grey squares) using Equation \eqref{eq:force0}. The three sums in Equation \eqref{eq:force0} are labeled with stars ($1^{st}$ sum), squares ($2^{nd}$ sum), and triangles ($3^{rd}$ sum).}
\label{fig:convolution}
\end{figure}

Here, we provide a simple numerical verification for the proposed mechanism in Section \ref{sec:force}, namely, the nonlinear driving terms in the convolution form in Fourier space favor the low azimuthal wavenumber modes. As an example, we consider the square of azimuthal velocity perturbation $v^2$ located near $r_2$ for $l=2$. In Figure \ref{fig:convolution}, we show the $l$-th Fourier amplitude $|(v^2)_l|$ by two methods, namely, direct computation by FFT and summation of linear amplitudes using Equation \eqref{eq:force0}. The previous numerical solution $v$ of $q=10^{-6}$ (i.e., first row of Figure \ref{fig:qcases}) is used as its relative perturbation smaller than 0.1\%. This ensures the nonlinear effect is minimal. For the latter method, we substitute $v$ into functions $f$ and $g$ on the right-hand side of Equation \eqref{eq:force0} and obtain three sums (also shown in Figure \ref{fig:convolution}). For a truncated Fourier series as used in numerical FFT, the second and third summations in Equation \eqref{eq:force0} differ only for large $l$. More importantly, these two terms (i.e., triangles and squares in Figure \ref{fig:convolution}) are responsible for the low-$l$ driving. Moreover, they correspond to the $m>l$ terms with ``wave-wave" or ``wave-non-wave" interactions described in Section \ref{sec:force} (see Figure \ref{fig:wave1}). On the other hand, the first summation in Equation \eqref{eq:force0} (i.e., $l>m$ terms) is responsible for the second peak at $l=11$ in Figure \ref{fig:convolution}. 

Finally, we note that our conclusions regarding the stronger driving of low azimuthal wavenumber modes do not depend on the numerical simulation itself as long as the shape of the Fourier spectrum is similar. The numerical result is used as input here because it provides a full solution of all linear waves that arrive at the particular UR. 

\section{Discussion}
\label{sec:Section7}

In this section, we discuss our results with application to understand previous findings based on numerical simulations.

\subsection{Existence of Secondary Spirals}

Using the numerical simulations, we find that the spiral wake or shock induced by a planet is generally double-peaked (Figure \ref{fig:qcases}). We note that the surface density perturbation for a low-mass planet ($q \leq 10^{-5}$) is less than 1\%. In any case, the left peaks in the left panels of Figure \ref{fig:qcases} moves away gradually when $q$ increases. As a result, the double-peaked spiral wave eventually becomes two separate spirals. Based on our numerical experiments, there is no lower limit for planet's mass for the secondary peak. In this case, the secondary peak is a genuine constructive interference feature due to linear waves, just like the primary spiral \citepalias{2002MNRAS.330..950O}. However, we note that 1\% perturbation in surface density is too small to be observed \citep[e.g.,][]{2015MNRAS.451.1147J}. 

On the other hand, the higher-order modes do not necessarily lead to more spiral arms. In Section \ref{sec:singlemode}, we demonstrate that the higher-order ultraharmonics do not produce additional spiral arms in the case of \emph{single} component of the planet's potential $\phi_m$ in the slightly nonlinear regime. Since the driving force is solely due to non-wave contribution (i.e., bar-like in \citetalias{1992ApJ...389..129A}), the higher-order waves are excited in-phase at their respective URs (c.f., Section \ref{sec:source}). At a particular radius, these waves do acquire some amount of phase difference during propagation. Without explicit viscosity, up to forth-order waves are found due to mode-coupling (see Figure \ref{fig5}). Therefore, a phase difference (i.e., non-zero $\Phi_{m,l}$) at excitation is likely to be a criteria for the secondary spirals to appear. 

One way for the higher-order waves to acquire a phase difference is to have wave-like driving force.  For single $\phi_m$, this is only possible for a third or higher-order wave, in which the interaction between $m$-th and $2m$-th waves generate a $m$-th wave (e.g., $q=10^{-3}$ case in Figure \ref{fig5}). The feedback due to resonant forcing will lead to higher-order changes in frequency and phase \citep{BenderOrszag1999}. For other processes, such as gravitational instability, this may result in nonlinear saturation \citep{1997ApJ...477..410L,2016arXiv160301280K}. We may leave the effects of resonant forcing in future investigations.

For a planet's potential, the wave-like contribution to $S_0$ is due to ``wave-wave" and ``wave-non-wave" interactions described in Section \ref{sec:force}. Since these two interactions primarily associated with the case with $l \leq m$ (Figure \ref{fig:wave1}), we speculate that the low-$l$ ultraharmonics wave may be indeed enhanced additionally by having more coherent phase. This can be seen by considering the relative phase of ultraharmonics wave and linear wave from Equation \eqref{eq:source6}, which is given by
\begin{align}
\label{eq:discuss0}
\Delta = \Phi_{m,l} - \frac{1}{3}(q^{(l)}_m)^3.
\end{align}
In the limit of $m \gg l$, the above expression becomes independent of $m$ (Equations \eqref{eq:source3} and \eqref{eq:source4}). The corresponding phase shift \citepalias[c.f.,][]{2002MNRAS.330..950O} for a $l$-th mode is $\Delta/l \simeq \pm 2/3h_0 + \mathcal{O}(l^{-1})$, which may facilitate the constructive interference among the second-order waves. Finally, as noted in \citet{2003ApJ...596..220C}, the importance of torque-cutoff effects may be important for high-$l$ ultraharmonics waves as well.

\subsection{Angular Offset of Secondary Spirals}

In general, secondary spirals appear as a distinct entity when $q$ is larger than $10^{-4}$ \citep{2006MNRAS.370..529D}. However, based on the mode-coupling, the number of arms is not directly related to the order of nonlinearity. For example, a third spiral arm can be a second-order effect instead of third-order, since such arm is not necessarily much weaker (see Figures \ref{fig0} and \ref{fig:qcases}).

Here we provide an alternative explanation. As shown in the previous sections, a planet's potential is capable of generating a number of spiral arms by nonlinear mode-coupling to the second-order. Therefore, without resorting to the higher-order modes, the second-order solution $u^{(2)}$ already contains multiple ultraharmonics waves with different number of arms. Indeed, the interplay between multiple harmonics is crucial to explain the offset between the apparent secondary spiral and the ``primary" spiral that traced to the planet.

In Section \ref{sec:modecouplelinear}, we demonstrate that the mode-coupling can result in a very different Fourier spectrum compared to the linear one (Figure \ref{fig:convolution}).
As a combination of the linear and higher-order modes, the relative Fourier amplitude between two components depends on the higher-order corrections, especially for the low-$l$ modes, which is presumably dominant. To understand such effect, we consider a simple wave solution that mimics the superposition of linear and ultraharmonics waves. The surface density perturbation in the inner disk can be expressed as
\begin{align}
\label{eq:discuss1}
\Delta\sigma(r,\varphi) = s_2(r) \cos(2\varphi+\delta_2) + s_3(r) \cos(3\varphi+\delta_3),
\end{align}
where $s_m$ and $\delta_m$ are the amplitude and phase of the $m$-th component of surface density response. For simplicity, we do not include the higher-order $m=0$ perturbation. In addition, we do not explicitly distinguish the linear and ultraharmonics waves, as they both contribute to the $m$-th harmonics. At a particular radius, the angular offset between the primary and secondary spirals is defined by the surface density peaks separation along azimuthal direction. Therefore, the angular offset equals to the difference of roots of $\varphi$-derivative of Equation \eqref{eq:discuss1}, which is a transcendental equation. In any case, the offset is basically determined by relative amplitude between $s_2$ and $s_3$. As a result, the offset may change from $2\pi/3$ to $\pi$ when the dominant Fourier mode in gas response changes from $m=3$ to $m=2$, which is in principle governed by the mass ratio. For example, the dominant Fourier mode for $q=10^{-5}$ is $m=6$ and the corresponding separation between spiral peaks is roughly $\pi/3$ (Figure \ref{fig:qcases}). Finally, we note that the torque cutoff and gap-opening effects, which reduce the torque by high-$m$ LRs that are close to the planet \citep{1980ApJ...241..425G,1993ApJ...419..155A}, may also play a role for the enhancement of $m=2$ spiral structure. However, we speculate this is more relevant to the case with a very wide gap, but a detailed discussion would be beyond the scope of the present work. In any case, in order to determine the amplitude and phase of each Fourier component, the details of the nonlinear driving terms are required. 

\subsection{Three-dimensional Disk Structure}

We expect several similarities and differences when a realistic three-dimensional (3D) disk is considered. As explored in the literature, the linear waves in 3D disk have different dispersion relations and propagate differently \citep{1993ApJ...409..360L,1998ApJ...504..983L,1999ApJ...515..767O,2002MNRAS.332..575B,2015ApJ...814...72L}. In particular, the boundary condition at the disk surface becomes important to determine which vertical modes are important \citep{1999ApJ...515..767O,2015ApJ...814...72L}. However, we speculate that the nonlinear mode-coupling can still occur as long as the wave forms between respective linear modes are similar (e.g., the fundamental modes can couple effectively with themselves). This mode-coupling can be understood by considering a quadratic term in the form $u_{m,n}(z)u_{m',n'}(z)$, where $m$ and $n$ are the quantum numbers for the azimuthal and vertical directions. As a result, the preferential excitation for low-$m$ modes can occur. Indeed, \citet{2015ApJ...813...88Z} and \citetalias{2015ApJ...815L..21F} found no substantial difference regarding the spiral structure (e.g., pitch angles) at the midplane in the 2D/3D simulations with idealized vertical temperature profiles, although vertical motion above the midplane appears. This may indicate the mode-coupling works the same way in 3D.

\subsection{Implications and Future Work}

Several quantitative analysis are left for further investigation. In particular, the corrections due to gas pressure \citep{1993ApJ...419..155A,1993ApJ...419..166A} in Equation \eqref{eq:ur_6} (i.e., $\gamma$ and $\epsilon$) and a treatment for resonant forcing \citep{1998ApJ...504..945L} are needed to provide a full understanding of how the Fourier spectra in Figure \ref{fig:qcases} depend on planet's mass. We speculate the relative amplitude of the most dominant modes govern the angular separations between primary and secondary spirals found in \citetalias{2015ApJ...815L..21F}. Also, the dependence on scale-height and temperature are not included in the current analysis. However, the pressure correction to the phase offset of ultraharmonics waves (i.e., Equation \eqref{eq:discuss0}) goes as $1/h_0$ which may result in insignificant change due to the modulo $2\pi$ property of the angular variable. For example, the approximate phase shifts are $\Delta/l = 2/3h_0\,(\mathrm{mod}\,2\pi) = 4.10$, $4.13$, $0.38$, and $4.17$ for $h_0=0.04$, $0.063$, $0.1$, and $0.16$, respectively, as adopted by \citetalias{2015ApJ...815L..21F}. Finally, the pressure correction may have an impact on the torque and propagation of ultraharmonics waves as well.

\section{Summary and Conclusion}
\label{sec:Section8}

By extending the linear theory of planet-disk interaction into slightly nonlinear regime, we investigate the gas response due to nonlinear mode-coupling. The excitation of higher-order waves is considered. In particular, the nonlinear driving terms are obtained for the second-order ultraharmonics, which is in a convolution form in Fourier space. This particular feature, along with the moderate phase coherence for high-$m$ harmonics, gives rise to the preferential excitation of waves with low azimuthmal wavenumber. The higher-order correction to the Fourier amplitudes of the gas response results in an angular offset between the peaks of the spiral waves. 

An analytical framework for the ultraharmonics waves is developed based on \citetalias{1992ApJ...389..129A}, which was in the context of spiral galaxies. The theory is extended to study a planet's potential that contains many Fourier components. Four kinds of mode-coupling are identified by considering the interactions between wave-like and non-wave response of the linear modes. For each harmonics of higher-order waves, there are two interactions associated mode-coupling (Equations \eqref{eq:source7a} and \eqref{eq:source7b}). We demonstrate that the ``wave-wave" interaction is responsible for the excitation for low-$m$ modes and its phase shift. 

Without attacking the Equation \eqref{eq:lh_7} directly, we numerically verify some analytical results regarding the phase and strength of higher-order modes using hydrodynamics simulations. We demonstrate that the mode-coupling, unlike other nonlinear effects, do not require large finite amplitude to occur. For example, in our inviscid simulation with a $m=3$ potential and $q=10^{-5}$, we are able to pick up the fourth-order wave (Figure \ref{fig5}). 

We apply this work to understand the multi-armed spiral structure in a protoplanetary disk. We present our attempts to understand the following features from previous numerical simulations \citet{2015ApJ...813...88Z} and \citetalias{2015ApJ...815L..21F}:
\begin{enumerate}
\item the existence of secondary spirals,
\item the angular separation between the primary and secondary spirals depends on the mass ratio $q$,
\item the separation becomes $\pi$ when the mass ratio is large, and
\item the secondary spiral is relative weaker in the outer disk outside the planet's orbit.
\end{enumerate}
Qualitative explanation for point 2 above is discussed in Section \ref{sec:Section7}, in which a more quantitative investigation may be needed. As a conclusion, the nonlinear mode-coupling is a promising mechanism to explain the multi-armed spiral structure induced by one planet. The formulation developed in this paper provides some insights and theoretical basis for future quantitative analysis on the observational signature regarding the spiral structure in a protoplanetary disk.  

\acknowledgements
Author appreciates the valuable comments and suggestions from Min-Kai Lin, Hsiang-Hsu Wang, and Frank Shu and thanks Pin-Gao Gu for stimulating discussions. Author is also grateful to Ron Taam for his help at Academia Sinica Institute of Astronomy and Astrophysics (ASIAA) and David C.C. Yen for providing an early version of {\it Antares} in cylindrical coordinates. Author would like to acknowledge the support from Academia Sincia Postdoctoral Fellowship and from the Theoretical Institute for Advanced Research in Astrophysics (TIARA) based in ASIAA.

\appendix
\section{Method of Solution}
\label{sec:methodofsolution}

In the limit of $\gamma \rightarrow 0$ and $\epsilon \rightarrow 0$, the governing equation for the second-order, $l$-armed ultraharmonics near the URs at $r=r_2$ is given by
\begin{align}
\label{eq:solve1}
\frac{d^2\sigma_l}{dX^2} - \frac{1}{X}\frac{d\sigma_l}{dX} - X \sigma_l = S_0(X),
\end{align}
where $X = (r-r_2)/(r_2 \Lambda)$ is a scaled distance to the UR and $S_0(X)$ is the nonlinear driving term. Expanding Equation \eqref{eq:lh_8c} in the neighborhood of $r_2$ and using Equations \eqref{eq:ur_4} and \eqref{eq:ur_7d}, we have
\begin{align}
\label{eq:solve2}
S_0(X) = -\frac{r_2\Lambda}{\Omega_l c^2}X\frac{d}{dr}\left[\frac{W_r(X)}{X}\right],
\end{align}
where $W_r(X)$ is the function $W_r$ in the proximity of $r_2$. We note that only $W_r$, which is associated with resonant denominator $D_l$, is relevant. As discussed in Section \ref{sec:force}, the forcing function $W_r$ can be expressed in a sum of two summation series (over $m$), $W_r(X) = W^+(X) + W^-(X)$, where the superscript indicates the interaction between $m$-th and $|m\pm l|$-th modes. Each of the summation series $W^\pm$ associates with the effective wavenumber $q^{(l)\pm}_{m}$ which is given by Equations \eqref{eq:source7a} and \eqref{eq:source7b}. Since Equation \eqref{eq:solve1} is correct to the linear order in $X$, we only need to evaluate the amplitude of $W^\pm$ to zeroth order.  Thus, the driving term reads
\begin{align}
\label{eq:solve3}
S_0(X) = \sum_{m,\pm} \left[ a^{(l)}_{m} X \frac{d}{dX}\left(\frac{e^{iq^{(l)}_{m} X}}{X} \right)\right],
\end{align}
where $a^{(l)}_{m}$ is the complex coefficient of the driving terms and we drop the plus and minus signs of $a^{(l)\pm}_{m}$ and $q^{(l)\pm}_{m}$ for clarity. The right-hand side of Equation \eqref{eq:solve3} is a double summation that sums over $m$ and over the $\pm$ signs for the two wave interactions.  The method presented in \citetalias{1992ApJ...389..129A} can still be applied here. To proceed, we begin with the two linear homogeneous solutions to Equation \eqref{eq:solve1} (c.f., \citealt{1984ApJ...281..600Y,1987Icar...69..157M}), namely, $Ai'(X)$ and $Bi'(X)$, which are the first-derivative of the Airy functions. The general solution is a sum of the homogeneous solutions and the particular solution, which is given by
\begin{align}
\label{eq:solve5}
\sigma_l(X) &= aAi'(X) + bBi'(X) - \pi Ai'(X)\int^X_{-\infty} Bi'(z)\frac{S_0(z)}{z} dz
+ \pi Bi'(X)\int^X_\infty Ai'(z)\frac{S_0(z)}{z} dz,
\end{align}
where $a$ and $b$ are constants to be determined by the boundary conditions. The lower limits in the integrals are chosen such that $\sigma_l$ is bounded at $X \rightarrow \pm \infty$ (as $Bi'$ blows up at $X \rightarrow \infty$). By requiring $\sigma_l$ remains finite for $X \rightarrow \infty$, we get $b=0$. For $X \rightarrow -\infty$, we apply the radiation boundary condition for trailing waves (i.e., positive wavenumber $k>0$). Note that $X < 0$ corresponds to $r > r_2^+$ and $r < r_2^-$ where $r_2^\pm = (1\pm 1/l)^{2/3}$. Thus, we have
\begin{align}
\label{eq:solve6}
a = i \pi \int^{\infty}_{-\infty} Ai'(z) \frac{S_0(z)}{z}dz = -i\pi \sum_{m} a^{(l)}_{m} \exp \left[-\frac{i}{3}(q^{(l)}_{m})^3\right],
\end{align}
where the last expression is obtained by using Equation \eqref{eq:solve2} and integrating by parts. The full solution can be expressed as
\begin{align}
\label{eq:solve9}
\sigma_l(X) = \sum_m a^{(l)}_{m} \pi \left[Ai'(X) \int^X_{-\infty} Bi(z) e^{iq^{(l)}_{m} z} dz - 
Bi'(X) \int^X_{\infty} Ai(z) e^{iq^{(l)}_{m} z} dz
- i Ai'(X) \int^\infty_{-\infty} Ai(z) e^{iq^{(l)}_m z} dz \right],
\end{align}
which is analogous to Equation (40) in \citetalias{1992ApJ...389..129A}. The asymptotic form of $\sigma$ for $X \rightarrow -\infty$ is given by
\begin{align}
\label{eq:solve10}
\sigma_l(X) = \sum_m a^{(l)}_{m} \pi \left[\int^\infty_{-\infty} Ai(z) e^{iq^{(l)}_m z} dz \right] H(X),
\end{align}
where $H(X) = Bi'(X) - iAi'(X)$ and the integral equals $\exp -\frac{i}{3}(q^{(l)}_m)^3$. The asymptotic form of $H(X)$ is 
\begin{align}
H(X) \sim \frac{(-X)^{1/4}}{\sqrt{\pi}} \exp i\left[ - \frac{2}{3}(-X)^{2/3} + \frac{\pi}{4}\right].
\end{align}
Finally, we have
\begin{align}
\label{eq:solve11}
\sigma_l \sim \sum_m a^{(l)}_{m} \sqrt{\pi} (-X)^{1/4} \exp i\left[-\frac{2}{3}(-X)^{3/2}+\frac{\pi}{4}-\frac{q^{(l)3}_m}{3}\right].
\end{align}
We note that $a^{(l)}_m$ is generally complex with a phase $\Phi_{m,l}(r_2)$ discussed in Section \ref{sec:force}.

\section{Numerical Code}
\label{sec:NumericalCode}

\begin{figure}[!htb]
\centering
\includegraphics[width=0.45\textwidth]{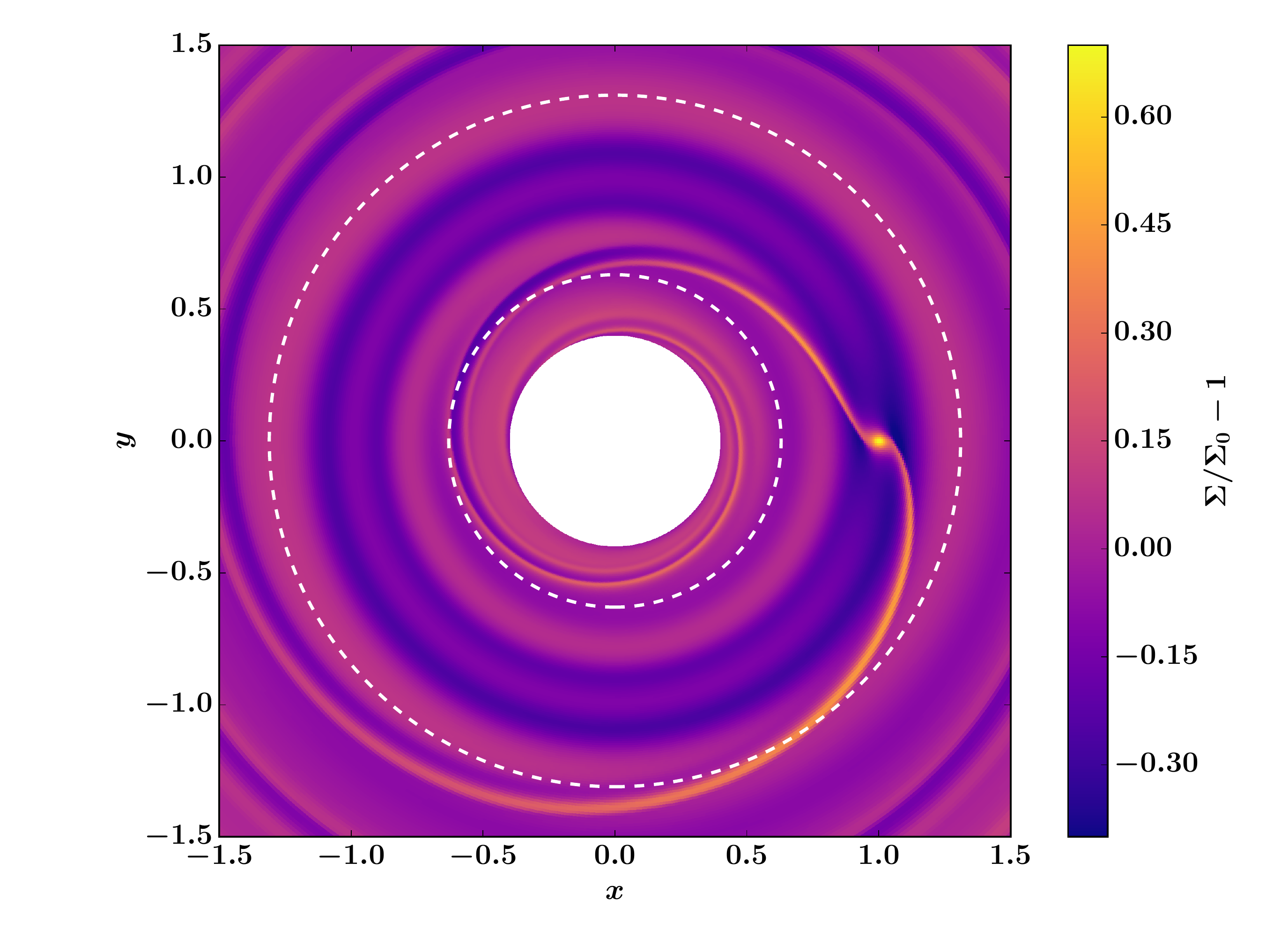}
\includegraphics[width=0.45\textwidth]{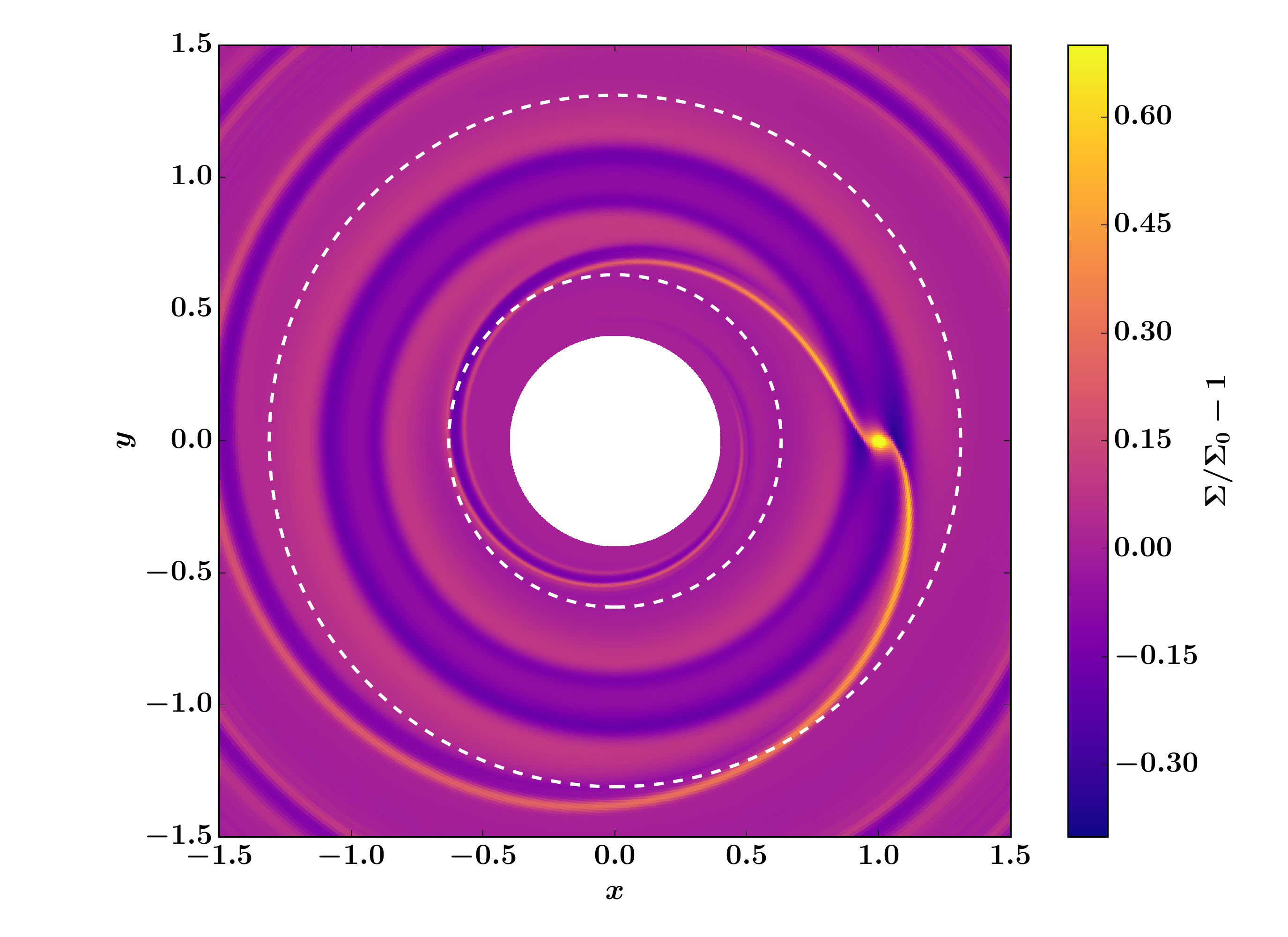}
\caption{A typical gas response due to a Neptune mass planet ($q=10^{-4}$). A single spiral density wake can be easily identified. The dashed white circles indicate the locations of LRs for $m=2$. The left and right panels show the results for our code and FARGO3D, respectively. The linear color scale shows the surface density perturbation $(\Sigma/\Sigma_0-1)$, which is truncated at $[-0.4,0.7]$.}
\label{fig10}
\end{figure}

A simple two-dimensional finite-volume hydrodynamics code is used to compute the numerical results in Section \ref{sec:Section6}. In this Appendix, we describe some more details. An exact Riemann solver for isothermal gas is used, along with piecewise linear method for interpolation. Non-reflective boundary condition are used at the inner and outer radii of the disk, where characteristic wave decompositions are performed \citep[c.f.,][]{Leveque2002}. Both self-gravity and viscosity are not included. A smoothing length is used for the planet's potential, which is chosen to be $r_s = 0.03 = 0.6 H_0$ for $H_0=0.05$. Since the numerical method of the higher-order Godunov code is involved (but well-known) and the planet-disk setup is fairly common in the community, we leave the discussion of numerical algorithm in a further code paper. Instead, we provide a comparison to other popular grid-based code FARGO3D \citep{2016ApJS..223...11B} in the two-dimensional setting for a Neptune-mass planet ($q=10^{-4}$) after about 50 orbits. The result is shown in Figure \ref{fig10}. Both simulations were run at the same resolution ($256\times768$) and with logarithmic grid spacing. The major difference is that the density contrast in the coorbital region, in which our code shows a deeper (partial) gap. This is probably due to different numerical viscosity present in the algorithm. We note that a wave-damping zone is used in the FARGO3D setup which occupies the inner-most cells of the disk and allows higher mass loss through the inner disk edge. In any case, the result is very similar and does not affect our conclusions on the spiral structure.

\bibliography{disk}

\begin{thebibliography}{}
\expandafter\ifx\csname natexlab\endcsname\relax\def\natexlab#1{#1}\fi

\bibitem[{{Artymowicz}(1993{\natexlab{a}})}]{1993ApJ...419..155A}
{Artymowicz}, P. 1993{\natexlab{a}}, \apj, 419, 155

\bibitem[{{Artymowicz}(1993{\natexlab{b}})}]{1993ApJ...419..166A}
---. 1993{\natexlab{b}}, \apj, 419, 166

\bibitem[{{Artymowicz} \& {Lubow}(1992)}]{1992ApJ...389..129A}
{Artymowicz}, P., \& {Lubow}, S.~H. 1992, \apj, 389, 129

\bibitem[{{Bate} {et~al.}(2002){Bate}, {Ogilvie}, {Lubow}, \&
  {Pringle}}]{2002MNRAS.332..575B}
{Bate}, M.~R., {Ogilvie}, G.~I., {Lubow}, S.~H., \& {Pringle}, J.~E. 2002,
  \mnras, 332, 575

\bibitem[{{Bender} \& {Orszag}(1999)}]{BenderOrszag1999}
{Bender}, C.~M., \& {Orszag}, S.~A. 1999, {Advanced Mathematical Methods for
  Scientists and Engineers} (Springer-Verlag, New York)

\bibitem[{{Benisty} {et~al.}(2015){Benisty}, {Juhasz}, {Boccaletti},
  {Avenhaus}, {Milli}, {Thalmann}, {Dominik}, {Pinilla}, {Buenzli}, {Pohl},
  {Beuzit}, {Birnstiel}, {de Boer}, {Bonnefoy}, {Chauvin}, {Christiaens},
  {Garufi}, {Grady}, {Henning}, {Huelamo}, {Isella}, {Langlois}, {M{\'e}nard},
  {Mouillet}, {Olofsson}, {Pantin}, {Pinte}, \& {Pueyo}}]{2015A&A...578L...6B}
{Benisty}, M., {Juhasz}, A., {Boccaletti}, A., {et~al.} 2015, \aap, 578, L6

\bibitem[{{Ben{\'{\i}}tez-Llambay} \& {Masset}(2016)}]{2016ApJS..223...11B}
{Ben{\'{\i}}tez-Llambay}, P., \& {Masset}, F.~S. 2016, \apjs, 223, 11

\bibitem[{{Chakrabarti} {et~al.}(2003){Chakrabarti}, {Laughlin}, \&
  {Shu}}]{2003ApJ...596..220C}
{Chakrabarti}, S., {Laughlin}, G., \& {Shu}, F.~H. 2003, \apj, 596, 220

\bibitem[{{Dai} {et~al.}(2015){Dai}, {Facchini}, {Clarke}, \&
  {Haworth}}]{2015MNRAS.449.1996D}
{Dai}, F., {Facchini}, S., {Clarke}, C.~J., \& {Haworth}, T.~J. 2015, \mnras,
  449, 1996

\bibitem[{{de Leon} {et~al.}(2015){de Leon}, {Takami}, {Karr}, {Hashimoto},
  {Kudo}, {Sitko}, {Mayama}, {Kusakabe}, {Akiyama}, {Liu}, {Usuda}, {Abe},
  {Brandner}, {Brandt}, {Carson}, {Currie}, {Egner}, {Feldt}, {Follette},
  {Grady}, {Goto}, {Guyon}, {Hayano}, {Hayashi}, {Hayashi}, {Henning},
  {Hodapp}, {Ishii}, {Iye}, {Janson}, {Kandori}, {Knapp}, {Kuzuhara}, {Kwon},
  {Matsuo}, {McElwain}, {Miyama}, {Morino}, {Moro-Martin}, {Nishimura}, {Pyo},
  {Serabyn}, {Suenaga}, {Suto}, {Suzuki}, {Takahashi}, {Takato}, {Terada},
  {Thalmann}, {Tomono}, {Turner}, {Watanabe}, {Wisniewski}, {Yamada}, {Takami},
  \& {Tamura}}]{2015ApJ...806L..10D}
{de Leon}, J., {Takami}, M., {Karr}, J.~L., {et~al.} 2015, \apjl, 806, L10

\bibitem[{{de Val-Borro} {et~al.}(2006){de Val-Borro}, {Edgar}, {Artymowicz},
  {Ciecielag}, {Cresswell}, {D'Angelo}, {Delgado-Donate}, {Dirksen}, {Fromang},
  {Gawryszczak}, {Klahr}, {Kley}, {Lyra}, {Masset}, {Mellema}, {Nelson},
  {Paardekooper}, {Peplinski}, {Pierens}, {Plewa}, {Rice}, {Sch{\"a}fer}, \&
  {Speith}}]{2006MNRAS.370..529D}
{de Val-Borro}, M., {Edgar}, R.~G., {Artymowicz}, P., {et~al.} 2006, \mnras,
  370, 529

\bibitem[{{Dong} {et~al.}(2015){Dong}, {Zhu}, {Rafikov}, \&
  {Stone}}]{2015ApJ...809L...5D}
{Dong}, R., {Zhu}, Z., {Rafikov}, R.~R., \& {Stone}, J.~M. 2015, \apjl, 809, L5

\bibitem[{{Fung} \& {Dong}(2015)}]{2015ApJ...815L..21F}
{Fung}, J., \& {Dong}, R. 2015, \apjl, 815, L21

\bibitem[{{Goldreich} \& {Tremaine}(1979)}]{1979ApJ...233..857G}
{Goldreich}, P., \& {Tremaine}, S. 1979, \apj, 233, 857

\bibitem[{{Goldreich} \& {Tremaine}(1980)}]{1980ApJ...241..425G}
---. 1980, \apj, 241, 425

\bibitem[{{Goodman} \& {Rafikov}(2001)}]{2001ApJ...552..793G}
{Goodman}, J., \& {Rafikov}, R.~R. 2001, \apj, 552, 793

\bibitem[{{Grady} {et~al.}(2013){Grady}, {Muto}, {Hashimoto}, {Fukagawa},
  {Currie}, {Biller}, {Thalmann}, {Sitko}, {Russell}, {Wisniewski}, {Dong},
  {Kwon}, {Sai}, {Hornbeck}, {Schneider}, {Hines}, {Moro Mart{\'{\i}}n},
  {Feldt}, {Henning}, {Pott}, {Bonnefoy}, {Bouwman}, {Lacour}, {Mueller},
  {Juh{\'a}sz}, {Crida}, {Chauvin}, {Andrews}, {Wilner}, {Kraus}, {Dahm},
  {Robitaille}, {Jang-Condell}, {Abe}, {Akiyama}, {Brandner}, {Brandt},
  {Carson}, {Egner}, {Follette}, {Goto}, {Guyon}, {Hayano}, {Hayashi},
  {Hayashi}, {Hodapp}, {Ishii}, {Iye}, {Janson}, {Kandori}, {Knapp}, {Kudo},
  {Kusakabe}, {Kuzuhara}, {Mayama}, {McElwain}, {Matsuo}, {Miyama}, {Morino},
  {Nishimura}, {Pyo}, {Serabyn}, {Suto}, {Suzuki}, {Takami}, {Takato},
  {Terada}, {Tomono}, {Turner}, {Watanabe}, {Yamada}, {Takami}, {Usuda}, \&
  {Tamura}}]{2013ApJ...762...48G}
{Grady}, C.~A., {Muto}, T., {Hashimoto}, J., {et~al.} 2013, \apj, 762, 48

\bibitem[{{Juh{\'a}sz} {et~al.}(2015){Juh{\'a}sz}, {Benisty}, {Pohl},
  {Dullemond}, {Dominik}, \& {Paardekooper}}]{2015MNRAS.451.1147J}
{Juh{\'a}sz}, A., {Benisty}, M., {Pohl}, A., {et~al.} 2015, \mnras, 451, 1147

\bibitem[{{Kley}(1999)}]{1999MNRAS.303..696K}
{Kley}, W. 1999, \mnras, 303, 696

\bibitem[{{Korycansky} \& {Pollack}(1993)}]{1993Icar..102..150K}
{Korycansky}, D.~G., \& {Pollack}, J.~B. 1993, \icarus, 102, 150

\bibitem[{{Kratter} \& {Lodato}(2016)}]{2016arXiv160301280K}
{Kratter}, K.~M., \& {Lodato}, G. 2016, ArXiv e-prints, arXiv:1603.01280

\bibitem[{{Laughlin} \& {Korchagin}(1996)}]{1996ApJ...460..855L}
{Laughlin}, G., \& {Korchagin}, V. 1996, \apj, 460, 855

\bibitem[{{Laughlin} {et~al.}(1997){Laughlin}, {Korchagin}, \&
  {Adams}}]{1997ApJ...477..410L}
{Laughlin}, G., {Korchagin}, V., \& {Adams}, F.~C. 1997, \apj, 477, 410

\bibitem[{{Laughlin} {et~al.}(1998){Laughlin}, {Korchagin}, \&
  {Adams}}]{1998ApJ...504..945L}
---. 1998, \apj, 504, 945

\bibitem[{{Lee} \& {Gu}(2015)}]{2015ApJ...814...72L}
{Lee}, W.-K., \& {Gu}, P.-G. 2015, \apj, 814, 72

\bibitem[{{Lesur} {et~al.}(2015){Lesur}, {Hennebelle}, \&
  {Fromang}}]{2015A&A...582L...9L}
{Lesur}, G., {Hennebelle}, P., \& {Fromang}, S. 2015, \aap, 582, L9

\bibitem[{LeVeque(2002)}]{Leveque2002}
LeVeque, R. 2002, Finite Volume Methods for Hyperbolic Problems, Cambridge
  Texts in Applied Mathematics (Cambridge University Press)

\bibitem[{{Lin} \& {Papaloizou}(1979)}]{1979MNRAS.186..799L}
{Lin}, D.~N.~C., \& {Papaloizou}, J. 1979, \mnras, 186, 799

\bibitem[{{Lin} \& {Papaloizou}(1993)}]{1993prpl.conf..749L}
{Lin}, D.~N.~C., \& {Papaloizou}, J.~C.~B. 1993, in Protostars and Planets III,
  ed. E.~H. {Levy} \& J.~I. {Lunine}, 749--835

\bibitem[{{Lin}(2015)}]{2015MNRAS.448.3806L}
{Lin}, M.-K. 2015, \mnras, 448, 3806

\bibitem[{{Lin} \& {Papaloizou}(2011)}]{2011MNRAS.415.1445L}
{Lin}, M.-K., \& {Papaloizou}, J.~C.~B. 2011, \mnras, 415, 1445

\bibitem[{{Lubow}(1990)}]{1990ApJ...362..395L}
{Lubow}, S.~H. 1990, \apj, 362, 395

\bibitem[{{Lubow}(1991)}]{1991ApJ...381..259L}
---. 1991, \apj, 381, 259

\bibitem[{{Lubow} \& {Ogilvie}(1998)}]{1998ApJ...504..983L}
{Lubow}, S.~H., \& {Ogilvie}, G.~I. 1998, \apj, 504, 983

\bibitem[{{Lubow} \& {Pringle}(1993)}]{1993ApJ...409..360L}
{Lubow}, S.~H., \& {Pringle}, J.~E. 1993, \apj, 409, 360

\bibitem[{{Lynden-Bell} \& {Kalnajs}(1972)}]{1972MNRAS.157....1L}
{Lynden-Bell}, D., \& {Kalnajs}, A.~J. 1972, \mnras, 157, 1

\bibitem[{{Meru}(2015)}]{2015MNRAS.454.2529M}
{Meru}, F. 2015, \mnras, 454, 2529

\bibitem[{{Meyer-Vernet} \& {Sicardy}(1987)}]{1987Icar...69..157M}
{Meyer-Vernet}, N., \& {Sicardy}, B. 1987, \icarus, 69, 157

\bibitem[{{Montesinos} {et~al.}(2016){Montesinos}, {Perez}, {Casassus},
  {Marino}, {Cuadra}, \& {Christiaens}}]{2016arXiv160107912M}
{Montesinos}, M., {Perez}, S., {Casassus}, S., {et~al.} 2016, ArXiv e-prints,
  arXiv:1601.07912

\bibitem[{{Muto} {et~al.}(2012){Muto}, {Grady}, {Hashimoto}, {Fukagawa},
  {Hornbeck}, {Sitko}, {Russell}, {Werren}, {Cur{\'e}}, {Currie}, {Ohashi},
  {Okamoto}, {Momose}, {Honda}, {Inutsuka}, {Takeuchi}, {Dong}, {Abe},
  {Brandner}, {Brandt}, {Carson}, {Egner}, {Feldt}, {Fukue}, {Goto}, {Guyon},
  {Hayano}, {Hayashi}, {Hayashi}, {Henning}, {Hodapp}, {Ishii}, {Iye},
  {Janson}, {Kandori}, {Knapp}, {Kudo}, {Kusakabe}, {Kuzuhara}, {Matsuo},
  {Mayama}, {McElwain}, {Miyama}, {Morino}, {Moro-Martin}, {Nishimura}, {Pyo},
  {Serabyn}, {Suto}, {Suzuki}, {Takami}, {Takato}, {Terada}, {Thalmann},
  {Tomono}, {Turner}, {Watanabe}, {Wisniewski}, {Yamada}, {Takami}, {Usuda}, \&
  {Tamura}}]{2012ApJ...748L..22M}
{Muto}, T., {Grady}, C.~A., {Hashimoto}, J., {et~al.} 2012, \apjl, 748, L22

\bibitem[{{Ogilvie} \& {Lubow}(1999)}]{1999ApJ...515..767O}
{Ogilvie}, G.~I., \& {Lubow}, S.~H. 1999, \apj, 515, 767

\bibitem[{{Ogilvie} \& {Lubow}(2002)}]{2002MNRAS.330..950O}
---. 2002, \mnras, 330, 950

\bibitem[{{Rafikov}(2002)}]{2002ApJ...572..566R}
{Rafikov}, R.~R. 2002, \apj, 572, 566

\bibitem[{{Rafikov}(2016)}]{2016arXiv160103009R}
---. 2016, ArXiv e-prints, arXiv:1601.03009

\bibitem[{{Shu}(2016)}]{ShuARAA2016}
{Shu}, F.~H. 2016, \araa, 54, in press

\bibitem[{Shu {et~al.}(1973)Shu, Milione, \& Roberts}]{SMR1973}
Shu, F.~H., Milione, V., \& Roberts, W.~W. 1973, \apj, 183, 819

\bibitem[{{Shu} {et~al.}(1990){Shu}, {Tremaine}, {Adams}, \&
  {Ruden}}]{1990ApJ...358..495S}
{Shu}, F.~H., {Tremaine}, S., {Adams}, F.~C., \& {Ruden}, S.~P. 1990, \apj,
  358, 495

\bibitem[{{Shu} {et~al.}(1985){Shu}, {Yuan}, \&
  {Lissauer}}]{1985ApJ...291..356S}
{Shu}, F.~H., {Yuan}, C., \& {Lissauer}, J.~J. 1985, \apj, 291, 356

\bibitem[{{Stolker} {et~al.}(2016){Stolker}, {Dominik}, {Avenhaus}, {Min}, {de
  Boer}, {Ginski}, {Schmid}, {Juhasz}, {Bazzon}, {Waters}, {Garufi},
  {Augereau}, {Benisty}, {Boccaletti}, {Henning}, {Maire}, {Menard}, {Meyer},
  {Langlois}, {Pinte}, {Quanz}, {Thalmann}, {Beuzit}, {Carbillet}, {Costille},
  {Dohlen}, {Feldt}, {Gisler}, {Mouillet}, {Pavlov}, {Perret}, {Petit},
  {Pragt}, {Rochat}, {Roelfsema}, {Salasnich}, {Soenke}, \&
  {Wildi}}]{2016arXiv160300481S}
{Stolker}, T., {Dominik}, C., {Avenhaus}, H., {et~al.} 2016, ArXiv e-prints,
  arXiv:1603.00481

\bibitem[{{Takami} {et~al.}(2014){Takami}, {Hasegawa}, {Muto}, {Gu}, {Dong},
  {Karr}, {Hashimoto}, {Kusakabe}, {Chapillon}, {Tang}, {Itoh}, {Carson},
  {Follette}, {Mayama}, {Sitko}, {Janson}, {Grady}, {Kudo}, {Akiyama}, {Kwon},
  {Takahashi}, {Suenaga}, {Abe}, {Brandner}, {Brandt}, {Currie}, {Egner},
  {Feldt}, {Guyon}, {Hayano}, {Hayashi}, {Hayashi}, {Henning}, {Hodapp},
  {Honda}, {Ishii}, {Iye}, {Kandori}, {Knapp}, {Kuzuhara}, {McElwain},
  {Matsuo}, {Miyama}, {Morino}, {Moro-Martin}, {Nishimura}, {Pyo}, {Serabyn},
  {Suto}, {Suzuki}, {Takato}, {Terada}, {Thalmann}, {Tomono}, {Turner},
  {Wisniewski}, {Watanabe}, {Yamada}, {Takami}, {Usuda}, \&
  {Tamura}}]{2014ApJ...795...71T}
{Takami}, M., {Hasegawa}, Y., {Muto}, T., {et~al.} 2014, \apj, 795, 71

\bibitem[{{Takeuchi} {et~al.}(1996){Takeuchi}, {Miyama}, \&
  {Lin}}]{1996ApJ...460..832T}
{Takeuchi}, T., {Miyama}, S.~M., \& {Lin}, D.~N.~C. 1996, \apj, 460, 832

\bibitem[{{Tanaka} {et~al.}(2002){Tanaka}, {Takeuchi}, \&
  {Ward}}]{2002ApJ...565.1257T}
{Tanaka}, H., {Takeuchi}, T., \& {Ward}, W.~R. 2002, \apj, 565, 1257

\bibitem[{{Terquem}(2003)}]{2003MNRAS.341.1157T}
{Terquem}, C.~E.~J.~M.~L.~J. 2003, \mnras, 341, 1157

\bibitem[{{Wagner} {et~al.}(2015){Wagner}, {Apai}, {Kasper}, \&
  {Robberto}}]{2015ApJ...813L...2W}
{Wagner}, K., {Apai}, D., {Kasper}, M., \& {Robberto}, M. 2015, \apjl, 813, L2

\bibitem[{{Wang} {et~al.}(2014){Wang}, {Bu}, {Shang}, \&
  {Gu}}]{2014ApJ...790...32W}
{Wang}, H.-H., {Bu}, D., {Shang}, H., \& {Gu}, P.-G. 2014, \apj, 790, 32

\bibitem[{{Wang} {et~al.}(2015){Wang}, {Lee}, {Taam}, {Feng}, \&
  {Lin}}]{2015ApJ...800..106W}
{Wang}, H.-H., {Lee}, W.-K., {Taam}, R.~E., {Feng}, C.-C., \& {Lin}, L.-H.
  2015, \apj, 800, 106

\bibitem[{{Ward}(1997)}]{1997Icar..126..261W}
{Ward}, W.~R. 1997, \icarus, 126, 261

\bibitem[{{Yuan}(1984)}]{1984ApJ...281..600Y}
{Yuan}, C. 1984, \apj, 281, 600

\bibitem[{{Yuan} \& {Cassen}(1994)}]{1994ApJ...437..338Y}
{Yuan}, C., \& {Cassen}, P. 1994, \apj, 437, 338

\bibitem[{{Yuan} \& {Cheng}(1991)}]{1991ApJ...376..104Y}
{Yuan}, C., \& {Cheng}, Y. 1991, \apj, 376, 104

\bibitem[{{Yuan} \& {Yen}(2005)}]{2005JKAS...38..197Y}
{Yuan}, C., \& {Yen}, D.~C.~C. 2005, Journal of Korean Astronomical Society,
  38, 197

\bibitem[{{Zhu} {et~al.}(2015){Zhu}, {Dong}, {Stone}, \&
  {Rafikov}}]{2015ApJ...813...88Z}
{Zhu}, Z., {Dong}, R., {Stone}, J.~M., \& {Rafikov}, R.~R. 2015, \apj, 813, 88

\end{thebibliography}
\end{document}